\documentclass[manuscript,screen]{acmart} 

\usepackage{todonotes}
\usepackage{subcaption}
\usepackage{booktabs,tabularx, multirow}
\usepackage{xltabular}
\usepackage{adjustbox}
\usepackage{paralist}
\usepackage{enumitem}
\usepackage{colortbl}
\usepackage{color,soul}

\AtBeginDocument{%
  }

\setcopyright{acmlicensed}
\copyrightyear{2018}
\acmYear{2018}
\acmDOI{XXXXXXX.XXXXXXX}
\acmConference[Conference acronym 'XX]{Make sure to enter the correct
  conference title from your rights confirmation email}{June 03--05,
  2018}{Woodstock, NY}
\acmISBN{978-1-4503-XXXX-X/2018/06}

\begin{document}

\title[What People See (and Miss) About Generative AI Risks]{What People See (and Miss) About Generative AI Risks:\\Perceptions of Failures, Risks, and Who Should Address Them}

\author{Megan Li}
\email{meganli@andrew.cmu.edu}
\author{Wendy Bickersteth}
\email{wbickers@andrew.cmu.edu}
\author{Ningjing Tang}
\email{ningjint@andrew.cmu.edu}
\author{Parv Kapoor}
\email{parvk@andrew.cmu.edu}
\author{Khinezin Win}\authornote{Work completed while at Carnegie Mellon University}
\email{kwin@andrew.cmu.edu}
\author{Peter Zhong}\authornote{Additional affiliation: Replit}
\email{yongzhon@andrew.cmu.edu}
\author{Jason I. Hong}
\email{jasonh@cs.cmu.edu}
\author{Lorrie Faith Cranor}
\email{lorrie@cmu.edu}
\author{Hoda Heidari}
\email{hheidari@andrew.cmu.edu}
\author{Hong Shen}
\email{hongs@andrew.cmu.edu}

\affiliation{%
  \institution{\\Carnegie Mellon University}
  \city{Pittsburgh}
  \state{Pennsylvania}
  \country{USA}
}

\authorsaddresses{}
\renewcommand{\shortauthors}{Li et al.}

\begin{abstract}
Despite growing concerns about the risks of Generative AI (GenAI), there is limited understanding of public perceptions of these risks and their associated \textit{failure modes}---defined as recurring patterns of sociotechnical breakdown across the GenAI lifecycle that contribute to risks of real-world harm. To address this gap, we present a survey instrument, validated with eight subject matter experts and deployed on a sample of 960 U.S.-based participants, to assess awareness and perceptions of GenAI's failure modes, their associated risks, and stakeholder responsibilities to address them. To support realism and content validity, our instrument is structured around scenarios grounded in publicly reported incidents and a taxonomy of GenAI's failure modes. Findings suggest that our instrument is (1) effective for assessing risk awareness and perceptions in a way that is grounded in people's current contexts of use, yet is extensible to new contexts that will inevitably arise; and (2) potentially useful for informing the design of AI literacy tools and interventions. We argue for AI literacy and governance approaches that align with how people encounter and reason about GenAI in everyday life.
\end{abstract}



\received{20 February 2007}
\received[revised]{12 March 2009}
\received[accepted]{5 June 2009}

\maketitle

\section{Introduction}

Generative AI (GenAI) systems are increasingly embedded in everyday life, prompting growing public concern about the risks they pose \cite{thorbecke2023ai,eliot2024generative}. To define risk, we adapt the definition provided in the AI Risk Management Framework proposed by the U.S. National Institute of Standards and Technology: ``the composite measure of an event’s probability of occurring and the magnitude or degree of the consequences of the corresponding event'' \cite{nist_rmf} and use risk to refer specifically to risk of \textit{harm}. These risks---and their underlying sociotechnical failure modes, defined as ``recurring patterns of sociotechnical breakdown across the GenAI lifecycle that contribute to real-world harms or risks of harm'' \cite{mapping_2025}---have been documented and taxonomized in a growing body of work across Human-Computer Interaction (HCI), Fairness, Accountability, and Transparency (FAccT), and Science and Technology Studies (STS) communities, \cite{mapping_2025, weidinger_sociotechnical_2023, hutiri_not_2024, dominguez_foundation_risks}. Yet, there is still little understanding about how members of the general public perceive these risks and their associated failure modes. As GenAI is deployed across highly variable domains, public perceptions of its limitations and risks are increasingly consequential. Having a calibrated understanding of GenAI's failure modes and risks is critical for responsible use and governance of the technology \cite{ai_literacy}. 

While prior work has investigated public perceptions of GenAI, it tends to focus narrowly on specific risks or applications 
such as privacy, job displacement, or loss of human connection \cite{ai_companions_perceptions, ortega_data_flows_genai, privacy_custom_GPT}. In doing so, it often overlooks other types of risks, including those posed by basic failures of functionality \cite{raji_fallacy}. To address this gap, we structure our work around a comprehensive taxonomy of GenAI \textbf{failure modes} \cite{mapping_2025}. In particular, we investigate perceptions of GenAI's failure modes and the risks they pose, both broadly and in specific contexts of use. Because GenAI is a general-purpose technology that is constantly applied in new and emerging domains, thereby creating new risks, we argue that using failure modes rather than specific risks as a framing device can offer more generalizable insights into awareness and perceptions of GenAI's limitations and risks.

Our work is framed by three research questions.
\begin{itemize}
    \item \textbf{RQ1:} What GenAI failure modes are people aware of and how likely do they think they are?
    \item \textbf{RQ2:} What risks are people aware of and do they associate them with particular failure modes? 
    \item \textbf{RQ3: }Who do people think should address the risks of GenAI? 
\end{itemize}



To answer these questions, we developed a scenario-based survey instrument grounded in real-world GenAI incidents and a taxonomy of GenAI failure modes developed by Li et al.  \cite{mapping_2025}. Our survey elicits general attitudes toward GenAI, awareness of how GenAI's failure modes can manifest and cause risks in specific contexts of use, and perceptions of which stakeholders should address the risks associated with GenAI. It is structured around 24 hypothetical scenarios, each of which we developed to illustrate how failure modes can manifest in a particular context. To establish expert consensus on the failure modes illustrated in each scenario, we recruited eight responsible AI experts to evaluate our scenarios.
We administered our survey to 960 U.S.-based participants via Prolific.

Our findings suggest that our survey instrument is an effective tool for mapping perceptions and understandings of GenAI's risks across contexts of use. For instance, while prior work on AI risk perceptions has suggested that many people are aware of only a narrow set of societal risks \cite{Oppenlaender_2023}, our participants surfaced a rich array of perceived risks across different contexts of use. Additionally, our empirical results indicate that participants are widely aware of failure modes that are observable in generated outputs, such as hallucinations, but are often unaware of those reflecting upstream design and release choices, such as developers' problematic data collection and use. This has implications for AI literacy: the gaps revealed by our survey suggest that people can be competent at using GenAI as a tool while being unaware of the potentially harmful ways in which GenAI systems are developed. 
This is particularly relevant because many participants called on not only government and industry actors but also users and the public to call attention to, mitigate, and address the risks of GenAI.

In this work, we introduce and validate a scenario-based survey instrument to assess awareness of GenAI failure modes and their associated risks. We provide one of the first empirical studies mapping people’s understanding of GenAI failures and risks in diverse, real-world contexts of use. We offer our instrument as a tool to probe where misunderstandings of GenAI's limitations lie and how they can materialize into risky patterns of misuse, ultimately informing literacy interventions and governance frameworks that are grounded in the needs of real people. 



\section{Related Work}
The following provides an overview of prior work mapping the risks and failure modes of AI, public perceptions of these risks, and the relevance of both these topics to perspectives on GenAI literacy. Throughout, we highlight work that focuses specifically on GenAI and its unique risk landscape.


\subsection{Risks, Harms, and Failure Modes of GenAI}


Shelby et al. define harms from algorithmic systems as ``adverse lived experiences resulting from a system’s deployment and operation in the world'' ~\cite{shelby}. Drawing on literature from HCI, FAccT, and STS communities, their work is grounded in the idea that algorithmic harms occur not in a vacuum, but are experienced by people and arise from interaction of technical system components with societal power dynamics \cite{shelby}. A growing body of literature identifies and structures these into taxonomies of harms; for example, Abercrombie et al. developed a taxonomy of AI, algorithmic, and automation harms with nine high-level categories including \textit{Autonomy}, \textit{Psychological}, and \textit{Societal and cultural} harms \cite{abercrombie}. Slattery et al. reviewed and synthesized existing taxonomies of AI risks to develop a comprehensive repository of AI risks, which they organize into both a \textit{domain} taxonomy and a \textit{causal} taxonomy \cite{slattery2025airiskrepositorycomprehensive}.

Building on these foundations, recent studies have focused on the unique risks posed by GenAI. 
These risks---from misinformation to deepfakes---are conceptualized as sociotechnical phenomena that can affect many stakeholders, including end users of GenAI, third-party individuals (e.g., the subject of a deepfake), communities (e.g., through GenAI's repetition of negative stereotypes), and society (e.g., the breakdown of democratic institutions) \cite{hutiri_not_2024, weidinger_sociotechnical_2023, dominguez_foundation_risks,mapping_2025, generative_ghosts_risks}. Li et al. formalize the idea that algorithmic harms arise through sociotechnical interactions by introducing the term \textit{sociotechnical failure modes}, defined as ``the ways in which a [technical] system, its creators or users, or the various societal structures interacting with it bring risk or harm to some stakeholders.'' Building upon prior work by Raji et al. and Slattery et al. \cite{raji_fallacy, slattery2025airiskrepositorycomprehensive}, they map both technical and human failure modes to real-world harms through a systematic analysis of publicly-reported GenAI incidents. They propose a taxonomy of GenAI failure modes with four high-level categories: \textit{design-related}, \textit{development- and evaluation-related}, \textit{release-related}, and \textit{use-related} failure modes \cite{mapping_2025}.

While these taxonomies offer a robust framework for analyzing GenAI risks and failures, most rely on expert or incident-level analyses. Recent work has highlighted the general public as a crucial stakeholder in AI risk identification and management 
\cite{participatory_ai_prioritization}. However, as Abercrombie et al. note, 
it's still unclear how the public perceives the AI harms and failures they encounter, 
in part because they may lack the requisite vocabulary and knowledge to report them \cite{abercrombie}. 



\subsection{Public Perceptions of AI Risks}
Public perceptions of AI risks play a critical role in the adoption, acceptance, and governance of the technology \cite{aiportrayals_matter}.  
Prior work in HCI and related fields has highlighted widespread public concerns about 
AI's societal impacts, including those that have already been observed, such as privacy harms, unfair outcomes, and over-reliance, along with those that have yet to be observed, such as rogue AI \cite{Kim09122024, contextualizing_perceptions_biases, kelley_aiperceptions, cave_public_responses_ai, Narayanan02012024, starke2021fairnessperceptionsalgorithmicdecisionmaking, lay_perceptions_discrimination, yurrita_fairness}. 
With the rapid adoption of GenAI across domains and sectors, there is a growing body of work exploring how people perceive its risks. 
Similar to previous work on perceptions of non-generative AI, this work often focuses on perceptions of societal risks like bias and privacy \cite{Oppenlaender_2023, Lee2024Public, emotion_ai, ai_companions_perceptions, privacy_custom_GPT, ortega_data_flows_genai}, with some work investigating GenAI-specific risks such as AI-generated non-consensual intimate imagery \cite{ncii_perceptions} and ``perceptual harms''---defined as ``harms caused to users when others perceive or suspect them of using AI'' \cite{perceptual_harms, rae_perceived}. Oppenlaender et al. studied people's perceptions of text-to-image systems and found that while they were generally aware of societal risks of the technology (e.g., job displacement), they were ``oblivious of the potential personal risks'' of the technology to individual users (e.g., legal ramifications for copyright infringement) \cite{Oppenlaender_2023}. 
Lee et al. studied public perceptions of the risks and benefits of ChatGPT and found that participants with more knowledge about ChatGPT perceive greater risks, while those who had used ChatGPT tended to recognize its benefits \cite{Lee2024Public}.


Some work explores perceptions of blame and stakeholder responsibility with regard to AI harms \cite{blaming_AI, who_AI_governance, lima2021human, lima2025public}. Lima et al. found that in incidents of algorithmic harm, people overwhelmingly attributed the blame for that harm to the developers or users of the system \cite{blaming_AI}. David et al. note that in the U.S., ``knowledge of public perceptions about the shared governance responsibilities of AI remains largely unclear because of legislative inaction.'' They conducted an online survey of U.S. adults to study how they perceived AI governance responsibilities and found that their participants generally assigned significantly more responsibility to governments and individuals than to corporations \cite{who_AI_governance}. 

A key gap in this literature is the tendency to assume that AI tools function correctly, which, as Raji et al. argue, is often a fallacy \cite{raji_fallacy}. By focusing on 
societal (e.g., reinforcement of biases, deterioration of social skills) risks, much work on AI risk perceptions presumes the functionality of AI tools. Recent work by Tolsdorf et al. \cite{safety_perceptions_genAI} attempts to address this gap; they propose a measurement instrument for the systematic assessment of perceived risks of GenAI chatbots based on the widely-cited taxonomy of GenAI risks developed by Weidinger et al. \cite{weidinger_sociotechnical_2023}. They found that most risks were perceived as unlikely, except for unhelpful outputs and misinformation. 

\subsection{Perspectives on GenAI Literacy} 
\label{sec:ai-literacy}


The concept of AI literacy originated in educational contexts, with prior work defining AI literacy as ``a set of competencies that enables individuals to critically evaluate AI technologies; communicate and collaborate effectively with AI; and use AI as a tool online, at home, and in the workplace'' \cite{ai_literacy}. Early research in this area mostly explored cultivating AI literacy in K-12 education settings \cite{touretzky, Chiu2024-yt}, focusing on basic computational thinking and understanding of AI capabilities. Building on these foundations, there is a growing line of work in HCI and responsible AI communities emphasizing the importance of cultivating AI literacy in the general public, centered on the idea that AI literacy enables users to maximize the benefits of AI while minimizing its risks \cite{edwards, Strauß2021, xie_ai_literacy}. 

Recent work increasingly recognizes the importance of understanding AI's \textit{limitations and risks} as a crucial component of AI literacy. For example, Long \& Magerko define 17 competencies that comprise AI literacy, including the ability to identify AI's strengths and weaknesses \cite{ai_literacy}. Building on this work, Annapureddy et al. define a model of \textit{Gen}AI literacy consisting of 12 competencies, including the ability to assess the outputs of GenAI (e.g., for factuality) \cite{annapureddy_generative_2025}. 
These works represent a shift toward \textit{critical} AI literacy, defined by Strauß as ``the ability to comprehend the core features of an AI system and its (in-)compatibility with its particular application contexts in a more complex sociotechnical reality'' \cite{Strauß2021}. Strauß argues that this kind of literacy is necessary to combat automation bias, or the ``general risk of uncritically accepting the outcome of an automated system'' \cite{Strauß2021}. The necessity of this type of literacy is further highlighted in work on AI risks and risk perceptions \cite{Oppenlaender_2023, mapping_2025}, and there are increasing efforts within HCI communities to create artifacts such as AI failure cards \cite{tang_failurecards} or interactive games \cite{game_genAI_literacy} to cultivate this literacy. Orthogonally, emerging work builds on foundations in digital education and informatics to argue that robust AI literacy is composed of three perspectives: the technological perspective (\textit{how does the technology work?}), the sociocultural perspective (\textit{what are the effects?}), and the user-oriented perspective (\textit{how do I use it?}) \cite{digital_education, dagstuhl_triangle, gu_ai_literacy}. These perspectives form the \textit{Dagstuhl triangle} \cite{dagstuhl_triangle}.
To support this growing field, several 
measurement instruments have been developed to assess AI literacy and identify gaps. 
Ng et al. \cite{ng_ai_literacy} reviewed existing work on AI literacy and identified a variety of constructs and tools that researchers use to evaluate AI literacy, including knowledge tests \cite{Kandlhofer-literacy, lin2021modeling, wan2020smileycluster, rodriguez2021evaluation}, self-reported measures \cite{chai2020extended, dai2020promoting, gong2020k}, and project-based assessments \cite{kaspersen2021machine, watkins2020cosmology}. 
However, Gu \& Ericson \cite{gu_ai_literacy} find several gaps in this body of work, particularly in promoting critical literacy surrounding AI tools. They also find that these instruments tend to focus on just one perspective of AI literacy (i.e., one vertex of the Dagstuhl triangle): instruments measuring key competencies to \textit{use} AI rarely incorporate sociocultural literacy perspectives and vice versa, even though the combination of these perspectives may be essential for responsible and safe use of these tools \cite{gu_ai_literacy}. 


In this work, we take a first step toward filling these gaps. We introduce a scenario-based survey instrument designed to assess people's awareness of GenAI failure modes and their associated risks. Rather than asking participants to evaluate abstract risks, we ground our instrument in realistic, narrative scenarios based on publicly reported GenAI incidents. This design choice---well-established in the fields of HCI and usable security and privacy \cite{lee_health_ai_risks, design_fiction, ncii_perceptions}---enables us to surface not just what risks people fear, but which types of failure modes they understand, overlook, or misattribute. By using failure modes as our framing device, we aim to capture not only users’ ability to recognize familiar patterns of risks, but also their preparedness to anticipate and mitigate risks in novel and evolving contexts of use. This is inspired by work in learning sciences, which demonstrates that understanding the underlying mechanisms of observable phenomena (i.e., the failure modes underlying observed harms or risks) is critical for reasoning about these phenomena in new contexts \cite{Bransford_Brown_Cocking_2000, categorization_physics}. In doing so, we offer a measurement instrument for perceptions of GenAI risks that is grounded in realistic contexts of today, yet is extensible to the expanding and evolving contexts that will inevitably arise.

\section{Methods}

To address our research questions, we conducted an online survey where $n = 960$ participants from the United States were randomly assigned to engage in a scenario representing one or more of 12 GenAI failure modes. We recruited eight trained responsible AI experts to establish consensus on the failure modes represented by each of our scenarios.

\subsection{Scenario Design}
\label{scenarios}
To avoid introducing a known GenAI tool (e.g., ChatGPT) as a source of bias, we asked participants to imagine ``a hypothetical Generative AI chatbot called Waso that you can interact with in a conversational way. Waso generates text and images in response to your queries.'' We altered this description for three scenarios that required participants to imagine a more narrowly-scoped tool. In one scenario, Waso generated legal advice; in another, Waso generated veterinary advice; and in another, Waso was renamed WasoGame and was an online choose-your-own-adventure game (these are scenario IDs 13, 14, and 22, respectively, in Table \ref{tab:scenarios}). 

We developed 24 scenarios, each representing how one of 12 GenAI failure modes, such as \textit{Hallucination, Data quality issues,} or \textit{Malicious use} (see Table \ref{tab:scenarios} or Appendix~\ref{app:failure-modes} for exhaustive list of failure modes),
could manifest in a real-world context of use. These failure modes are identified and defined by Li et al. \cite{mapping_2025}, who also surfaced two other failure modes we excluded. These were \textit{Resource demands}, which we omitted because it does not tend to manifest in single incidents of user interaction with GenAI, and \textit{Lack of robustness against adversarial prompting}, which we omitted because of its redundancy with other failure modes (e.g., \textit{Malicious use, Failure of safety guardrails}). Li et al. \cite{mapping_2025} provide specific, occasionally non-intuitive definitions for each failure mode; for instance, the failure mode \textit{Data quality issues} is defined as ``Data reflects or exacerbates societal biases, or overrepresents a certain type of content, or is polluted with poor quality information which is not sufficiently cleaned.'' We provide these definitions in Appendix~\ref{app:failure-modes}.

{\footnotesize
\linespread{1}\selectfont\centering
\begin{xltabular}{\textwidth}{p{.2cm}|X|X}
\caption[GenAI User Study: Scenario design]{We developed 24 scenarios. Each scenario represents one of 12 failure modes and is composed of a \textit{use case} and \textit{failure} involving Waso/WasoGame, a fictional GenAI tool. We assign each scenario an ID number to be referred to in the text and figures. Scenarios 6 and 8 were re-categorized to reflect expert consensus, and additional plausible failure modes are indicated for scenarios 13, 21, 22, and 23 as described in Section \ref{validation-method}.} 
\Description{This table presents 24 scenarios organized by failure modes for a fictional GenAI tool called Waso/WasoGame. The table has three columns: ID (scenario numbers 1-24), Use case (context description), and Failure (what goes wrong).
The scenarios are grouped into 12 categories with gray header rows:
Hallucination (scenarios 1-2): False biographical information and citing non-existent laws.
Failure of commonsense reasoning (scenarios 3-4): Missing obvious facts like Kenya starting with "K" and suggesting nut-containing recipes for nut allergies.
Failure of safety guardrails (scenarios 5-6): Ignoring user corrections about gender and overly restrictive content policies.
Failure to comply with contextual norms (scenarios 7-8): Gibberish responses and reproducing paywalled content.
Problematic data collection and use (scenarios 9-10): Sharing personal information and revealing undisclosed user knowledge.
Data quality issues (scenarios 11-12): Generating racist stereotypes and changing pronouns inappropriately.
Lack of transparency about capabilities/limitations (scenarios 13-14): Poor legal advice with fake precedents and extreme medical recommendations.
Undisclosed/unwelcome use (scenarios 15-16): AI-generated newspaper submissions and student essays violating originality requirements.
Malicious use (scenarios 17-18): Generating conspiracy theories and misinformation about fake accidents.
Improper use (scenarios 19-20): Users not reviewing AI outputs before professional use.
Unanticipated use (scenarios 21-22): Using AI to detect AI-generated content and generating inappropriate gaming content.
Contested use (scenarios 23-24): Disputes over deepfakes and AI-generated art submissions.
Many scenarios reference real incidents with codes like "AIAAIC1356" and "AIID497."}
\label{tab:scenarios} \\

\hline \textbf{ID} & {\textbf{Use case}} & {\textbf{Failure}} \\ \hline 
\endfirsthead

\multicolumn{3}{c}%
{\tablename\ \thetable{} -- continued from previous page} \\
\hline \textbf{ID} & {\textbf{Use case}} & {\textbf{Failure}} \\ \hline 
\endhead

\hline \multicolumn{3}{|r|}{{Continued on next page}} \\ \hline
\endfoot

\hline
\endlastfoot

\multicolumn{3}{c}{\cellcolor[gray]{0.85}\textit{Hallucination}} \\ \hline
\multirow{2}{*}{1} & Suppose you are a college student taking a class. You ask Waso for a short bio of a public figure for a paper you are writing for this class. & Waso responds with inaccurate information about the public figure. \\ \hline
\multirow{2}{*}{2} & Suppose you’re involved in a disagreement with a neighbor over your property lines, so you ask Waso for advice about your legal rights. & Waso cites a nonexistent local law claiming you are legally entitled to move your fence boundary by two feet. \\ \hline

\multicolumn{3}{c}{\cellcolor[gray]{0.85}\textit{Failure of commonsense reasoning}} \\ \hline
\multirow{4}{*}{3} & Suppose you are running a trivia night and are creating a list of trivia questions. You ask Waso for a list of all the African countries that start with the letter  ``K.'' & When you prompt Waso to list all the countries that start with a ``K,'' it says that no countries start with that letter. (Note that this is wrong, since the country Kenya is located in Africa and its name starts with the letter ``K''). \textit{(based on incident AIID609)}\\ \hline
\multirow{2}{*}{4} & Suppose you are cooking dinner for a friend who’s allergic to nuts and ask Waso for nut-free recipe ideas. & Waso responds with a recipe that includes a nut-containing cereal. Waso does not warn you about this. \\ \hline

\multicolumn{3}{c}{\cellcolor[gray]{0.85}\textit{Failure of safety guardrails}} \\ \hline
\multirow{2}{*}{5} & Suppose you are an author. You use Waso to generate a story involving child characters. & Waso refuses your request, explaining that it will not generate any fictional content involving minors. \\ \hline

\multicolumn{3}{c}{\cellcolor[gray]{0.85}\textit{Failure to comply with contextual norms}} \\ \hline
\multirow{3}{*}{6} & Suppose you are a college student taking a writing class. You ask Waso to generate a character description for a male doctor for a creative writing assignment. & Waso consistently produces descriptions of a female doctor, despite your repeated attempts to correct the description to male. \\ \hline
\multirow{3}{*}{7} & Suppose you are a college student. You use Waso to summarize various articles you were assigned to read for a homework assignment.  & Waso replies with gibberish, stating, ``This paper is not a paper is not a paper is not a paper. Muchas gracias for your understanding.'' \textit{(based on incident AIAAIC1356)} \\ \hline

\multicolumn{3}{c}{\cellcolor[gray]{0.85}\textit{Problematic data collection and use}} \\ \hline
\multirow{2}{*}{8} & Suppose you are a college student and you use Waso to research a current event. & Waso responds with long excerpts from a paywalled news article. \textit{(based on incident AIAAIC1234)} \\ \hline
\multirow{2}{*}{9} & Suppose you are feeling bored and you use Waso to look up the addresses of nearby fun places to visit.  & Waso responds with an address, along with a name, email address, and phone number of a real person. \textit{(based on incident AIAAIC1209)}  \\ \hline
\multirow{3}{*}{10} & Suppose you ask Waso to help you brainstorm plans for an upcoming long weekend. & Waso responds, ``How about a trip? You haven’t been on a trip for months.'' While this is true, you haven’t mentioned this fact to Waso recently. \\ \hline

\multicolumn{3}{c}{\cellcolor[gray]{0.85}\textit{Data quality issues
}} \\ \hline
\multirow{2}{*}{11} & Suppose you are using Waso to generate character descriptions for a creative writing assignment.  & Waso generates descriptions that reflect racist stereotypes.  \\ \hline
\multirow{2}{*}{12} & Suppose you are a journalist. You use Waso to proofread a news article you have written about a female tech CEO. & Waso changes all of the pronouns in the article to he/him/his rather than she/her/hers. \\ \hline

\multicolumn{3}{c}{\cellcolor[gray]{0.85}\textit{Lack of transparency about model's capabilities/limitations
}} \\ \hline
\multirow{3}{*}{13} & Suppose you receive a parking ticket. You use Waso to help you draft legal documents to contest the ticket. & The legal demand letter generated by Waso cites nonexistent precedents and contains poor reasoning. \textit{(based on incident AIID497)} \\ \cline{2-3} 
& \multicolumn{2}{l}{\cellcolor{yellow!25}Other Plausible Failure Mode(s): \textit{Hallucination}}
\\ \hline
\multirow{3}{*}{14} & Suppose your dog is having diarrhea. You consult Waso to figure out what to do. & Waso responds, ``Your dog is at the end of its life. I recommend euthanasia.'' Waso also provides a list of nearby clinics that will perform euthanasia. \textit{(based on incident AIAAIC1751)} \\ \hline

\multicolumn{3}{c}{\cellcolor[gray]{0.85}\textit{Undisclosed/unwelcome use}} \\ \hline
\multirow{3}{*}{15} & Suppose that you want to submit an opinion piece to your local newspaper. The instructions provided by the newspaper require the opinion piece to be based on the author’s experience and/or expertise. & After Waso and similar chatbots are released, the newspaper receives submissions that include AI-generated content. \textit{(based on incident AIID562)} \\ \hline
\multirow{3}{*}{16} & Suppose that you are a high school student. You are assigned an essay in one of your classes; the instructor requires that students submit their own work. & Students in the class use Waso to partially or entirely generate their essays. \\ \hline

\multicolumn{3}{c}{\cellcolor[gray]{0.85}\textit{Malicious use}} \\ \hline
\multirow{3}{*}{17} & Suppose that you are a high school student. For one of your courses, the teacher assigns an in-class activity that includes using Waso to generate short stories. & Rather than engaging in the activity, some of the students use Waso to generate QAnon-style conspiracy theories. \textit{(based on incident AIAAIC0636)} \\ \hline
\multirow{3}{*}{18} & Suppose that you are a writer at a newspaper, and your editor allows the use of Waso for writing assistance. & A fellow writer uses Waso to generate convincing misinformation, including that there was a fatal train accident nearby. \textit{(based on incident AIAAIC1009)} \\ \hline

\multicolumn{3}{c}{\cellcolor[gray]{0.85}\textit{Improper use
}} \\ \hline
\multirow{2}{*}{19} & Suppose that you are a lawyer and your client is going through a divorce. You use Waso to help with your legal research. & You do not check over Waso’s outputs and simply copy them into the filings. \textit{(based on incident AIAAIC1371)} \\ \hline
\multirow{2}{*}{20} & Suppose you are taking a poetry class and the teacher uses Waso to provide feedback to students on their poems. & The teacher does not read over the Waso-generated feedback before sending it to the students. \textit{(based on incident AIAAIC1202)} \\ \hline

\multicolumn{3}{c}{\cellcolor[gray]{0.85}\textit{Unanticipated use
}} \\ \hline
\multirow{3}{*}{21} & Suppose that you are an instructor at a college, and suspect that many of the essays that your students have submitted were AI-generated. & You consult Waso to verify your suspicions by uploading your students’ essays and asking questions about them. \\ \cline{2-3} 
& \multicolumn{2}{l}{\cellcolor{yellow!25}Other Plausible Failure Mode(s): \textit{Contested use}} 
\\ \hline
\multirow{4}{*}{22} & Suppose that you are an avid player of WasoGame and participate in online forums where people share their WasoGame adventures. & People in these forums begin sharing strategies to get the game to generate stories depicting sexual encounters involving children. \textit{(based on incident AIID402)} \\ \cline{2-3} 
& \multicolumn{2}{l}{\cellcolor{yellow!25}Other Plausible Failure Mode(s): \textit{Failure of safety guardrails, Malicious use}}
\\ \hline

\multicolumn{3}{c}{\cellcolor[gray]{0.85}\textit{Contested use
}} \\ \hline
\multirow{3}{*}{23} & Suppose that while watching the nightly news, you see that images of your favorite celebrity engaging in illicit activities have surfaced. & The celebrity claims that these images are deepfakes and denies having ever engaged in such activities. \textit{(based on incident AIAAIC0234)}
\\ \cline{2-3} 
& \multicolumn{2}{l}{\cellcolor{yellow!25}Other Plausible Failure Mode(s): \textit{Malicious use}}
\\ \hline
\multirow{2}{*}{24} & Suppose that you're submitting a piece you created to an art contest. The contest rules require that your work be entirely original. & Your submission is flagged as AI-generated because it seems similar to a copyrighted work.  \\ 
\end{xltabular}}

We provide our scenarios in Table~\ref{tab:scenarios}. 
We designed these as stylized fictionalizations of real-world GenAI incidents, many of which we sourced from AI incident repositories \cite{aiid, aiaaic_repo}, indicated in Table~\ref{tab:scenarios}. Other incidents were recounted to us as personal anecdotes. 

\subsection{Scenario Validation}
\label{validation-method}
We recruited eight responsible AI experts to establish expert consensus on the failure modes represented by each of our scenarios. None of these experts participated in the earlier scenario development process. Further details on our procedure and the demographics of these experts are available in Section~\ref{app:val-details} in the appendices.

We split our set of 24 scenarios into two disjoint sets of 12 and presented each expert with one set. As such, each scenario was seen by four experts. For each scenario, we asked the expert to select the GenAI failure modes (presented alongside their definitions in Li et al. \cite{mapping_2025}) they thought best explained the scenario's outcome(s). They were then presented with a rationale, authored by the research team, explaining how a particular failure mode---the one that the scenario was initially designed to represent---could explain the scenario's outcome(s) and asked to rate their agreement with the rationale. Finally, we asked them to compare their initial selections to the one proposed in the rationale.

For each scenario where at least three out of four experts agreed with the presented rationale (21 out of 24 scenarios), we considered the scenario to be \textit{validated} in truly representing the failure mode it was designed to capture. If a different failure mode was indicated in three or more of the experts' initial selections (Scenarios 13, 21, 22, and 23), we include it as an additional plausible failure mode in Table~\ref{tab:scenarios}. For each scenario where two or more experts disagreed with the presented rationale (Scenarios 3, 6 and 8), we elevated the scenario to group discussion amongst the authors. If a different failure mode was indicated in three or more of the experts' initial selections (Scenarios 6 and 8), we re-categorized the scenario to that failure mode. Otherwise, the research team discussed the experts' free responses to reach consensus, as documented in Table~\ref{tab:scenarios} (Scenario 3).

\subsection{Survey Design}

In our online survey, each scenario was presented in two formats: a \textit{use-case-only} format, composed of only the use case description, and a \textit{use-case-failure} format, composed of the use case and failure descriptions concatenated together. Each participant was randomly assigned with equal probability to one of our 24 scenarios in one of these two formats, for a total of 48 study conditions (12 failure modes $\times$ two scenarios developed per failure mode $\times$ two scenario formats $=$ 48 total conditions).  The survey began with questions about participants' use of, sentiment toward, and previous experiences with GenAI. The next section introduced the hypothetical GenAI tool Waso/WasoGame and the scenario description. Participants then answered several questions about the scenario. 

We first asked participants whether they had encountered situations similar to the one described in the scenario. Then,
if the participant was shown the use-case-only format, we asked them to describe something they thought could ``go wrong involving GenAI in a scenario like this''---eliciting their awareness of possible failures---and how likely they thought this event would be. If the participant was shown the use-case-failure format, we instead asked them how likely they thought ``this hypothetical scenario or something similar is to happen in the real world,'' eliciting their perception of the failure presented. In both formats, we then asked them to describe a possible consequence of the failure (imagined by the participant in the former, described in the scenario in the latter)---eliciting their awareness of possible harms---and how likely they thought this consequence was.
In the final section, we asked participants who they felt should address the potential harms arising from GenAI. Our complete survey design is available in Appendix \ref{app:survey-design}.


\subsection{Data Collection \& Analysis}

\subsubsection{Participant Recruitment}
\label{sec:participants}
We recruited 960 participants between August 7 and August 13, 2025 using the Prolific crowdworking platform. We used Prolific's built-in representative sampling tool to recruit a U.S.-representative sample in terms of age, sex, and ethnicity.\footnote{See https://researcher-help.prolific.com/en/article/95c345 for details on Prolific's representative sampling feature.} All participants were over 18 and based in the U.S. at the time of our survey. Participants took a median of 7 minutes and 16 seconds to complete the survey and were compensated \$1.82, for an effective compensation rate of \$15.03/hour. This study was approved by our university's Institutional Review Board. 

During data analysis, we discovered that participants were shown an incorrect description of one of our scenarios (Scenario 14, use-case-failure format). On September 1, 2025, we recruited 20 new participants on Prolific; all were assigned to this scenario (with the correct description). Participants took a median of 6 minutes and 52 seconds to complete the survey and were compensated \$1.82 for an effective rate of \$15.90/hour. We removed the data from the participants we had originally recruited for this scenario ($n = 20$) and replaced it with this newly collected data.

To prevent our sample from being biased toward participants who have previous negative experiences with or opinions of GenAI, our recruitment message presented the study as an investigation of ``perceptions of and experiences with GenAI tools such as ChatGPT,'' but did not specifically mention failures or harms. We also took several measures to ensure data quality and ethical data collection, which are detailed in Appendices \ref{app:data-quality} and \ref{app:ethical-considerations}, respectively. 

\subsubsection{Data Analysis}
\label{data-analysis}

We performed exploratory data analysis and descriptive statistics on participants' responses to multiple choice and Likert scale questions.
We also conducted thematic analysis of participants' responses to six free-response questions about their opinions of GenAI (Q10, Q12), their awareness of GenAI failures and harms (Q16, Q18, Q22), and which entities they felt should address the potential harms of GenAI (Q24). 
Our complete codebooks are available in Appendices \ref{app:failure-modes}, \ref{app:risk-defs}, \ref{app:stakeholders}, and \ref{app:opinions}. Details on our procedure for thematic analysis are available in Appendix \ref{app:thematic-analysis}.

\subsection{Limitations} \label{limitations}

\subsubsection{Sample limitations} First, our sample is entirely US-based, which limits insights for non-US contexts. Second, although we initially recruited a representative sample from the US in terms of age, sex, and ethnicity, there was an error (noted in Section \ref{sec:participants}) in one of our scenarios that required additional data collection and replacement of $n = 20$ participants' data. Thus, we cannot make representativeness claims about our final data. We also found that our survey participants were unlikely to be nationally representative regarding familiarity with GenAI. A recent study by the Pew Research Center found that 34\% of U.S. adults have used ChatGPT \cite{us_chatgpt_use}, while 85\% (819 out of 960) of our participants reported using ChatGPT and 93\% (897 out of 960) reported using at least one GenAI tool. We conjecture that this is due to two factors: first, our recruitment call on Prolific was titled ``Generative AI Perceptions and Attitudes,'' likely biasing our sample toward participants with higher familiarity with and interest in GenAI; and second, prior work has found that Prolific participants tend to have more technology knowledge than the general public \cite{prolific_techsavvy}. 
Finally, we note that the use of GenAI to complete surveys is an endemic problem on Prolific and similar platforms, posing risks to the integrity of data collected through them \cite{zhang2025detectingusegenerativeai, Prolific_LLM_mitigation}. To mitigate these risks, we used several strategies (documented in detail in Appendix~\ref{app:data-quality}) shown to be highly effective to reduce and detect the use of GenAI \cite{Prolific_LLM_mitigation, wang_promptinjection}. Nonetheless, these strategies are not foolproof, and we acknowledge this as a limitation of our work. 

In summary, this work describes an online, U.S.-based convenience sample with high GenAI familiarity that serves as an initial step to validate our survey instrument. Future work may aim to recruit a more representative sample in terms of GenAI familiarity by omitting references to GenAI in recruitment materials and/or avoiding online crowdsourcing platforms; as well as deploy our survey with a probability-based or panel-weighted national sample for population-level insights.

\subsubsection{Design limitations} We designed each of our scenarios to represent a specific failure mode with the goal of achieving even coverage of the 12 GenAI failure modes we selected for this study. However, as noted in Section~\ref{validation-method}, expert consensus on which failure mode(s) are represented by each scenario was nontrivial, and there were several scenarios that plausibly represented more than one failure mode. \textit{Unanticipated use} emerged as particularly nuanced, with one of our experts noting, ``I have been thinking of waso as just a general purpose llm...in that case, I am not fully sure when we can claim that a certain use case is `unanticipated'.'' Thus, our scenarios representing this failure mode were contentious, reflecting a reality of GenAI: without specified use cases, its risk surface is extremely difficult to anticipate. 
Additionally, most of our scenarios ask participants to imagine themselves as users of Waso. This design choice may have discouraged them from imagining the users (themselves) as vectors of harm, thus making them less likely to surface use-related failure modes such as malicious use.

\section{Results}  \label{results}

In this section, we report the results of our online survey. We first summarize participants' prior experiences with and attitudes toward GenAI; we then present our results with respect to each of our three research questions.
Throughout this section, we include participants' responses as-is, without correcting their grammatical or spelling errors. 

\subsection{Prior Experiences With and Attitudes Toward GenAI}
\label{sec:attitudes}

Our participants consisted of 93\% GenAI users (i.e., they indicated using at least one GenAI tool) and 6.5\% GenAI non-users (i.e., they indicated that they did not use any GenAI tools). Of the GenAI users, most (57.4\%) considered themselves ``proficient users,'' defined in our survey as those who use ``Generative AI in advanced, efficient, and often creative ways to complete complex tasks and increase productivity.'' We refer to the 42.6\% of users who did not consider themselves ``proficient'' as \textit{casual} users. Further details on our participants' demographics and usage characteristics are available in Section~\ref{app:demographics-and-usage} in the appendices.

Generally, attitudes toward GenAI's capabilities were more positive than opinions about GenAI's impact. Opinions trended upward with level of GenAI use, with most non-users reporting negative opinions of both GenAI's capabilities and its impact, while both casual and proficient users overwhelmingly reported positive opinions about both.
Participants' reported experiences with GenAI offer one explanation for this difference: users were much more likely than non-users to report a previous positive experience with GenAI, and non-users were (unsurprisingly) unlikely to report any previous experience at all. This echoes previous work, which finds that more frequent users of GenAI (which our proficient users tended to be) had higher trust and perceived less risk associated with the technology \cite{safety_perceptions_genAI}. 
Figure~\ref{fig:opinions-and-experiences} in the appendices summarizes our participants' reported opinions and previous experiences with GenAI. 

\subsection{RQ1: Awareness and Perceptions of GenAI Failure Modes Likely Reflect Lived Experiences} \label{sec:rq1}

We used Li et al.'s taxonomy of failure modes \cite{mapping_2025} as a starting point to build our codebook and added the failure modes \textit{Over-reliance}, \textit{Poor/unsatisfactory output}, \textit{Parroting}, and \textit{Erroneous prompting} as codes that emerged during analysis. See Section~\ref{app:codebook-revisions} in the appendices for definitions. We found that most participants demonstrated awareness of development- and evaluation-related failure modes, which can largely be detected through examination of generated outputs. 

\subsubsection{Broad awareness of erroneous outputs, but limited awareness of upstream failure modes}
Throughout our survey, participants most often demonstrated awareness that GenAI outputs can be erroneous (e.g., through hallucinations or ``obvious clerical errors'') or otherwise unsatisfactory (e.g., by ``sounding robotic'' or demonstrating bias). By contrast, there were several failure modes that participants rarely described, despite our scenarios having been designed with these in mind. These tended to be failure modes related to upstream design choices or model release and regulation dynamics. \textbf{Overall, we found that participants' awareness or attention toward different failure modes likely reflected their direct/lived experiences with evaluating generated outputs.}

\begin{figure*}[t]
    \centering
        \includegraphics[width=.68\textwidth]{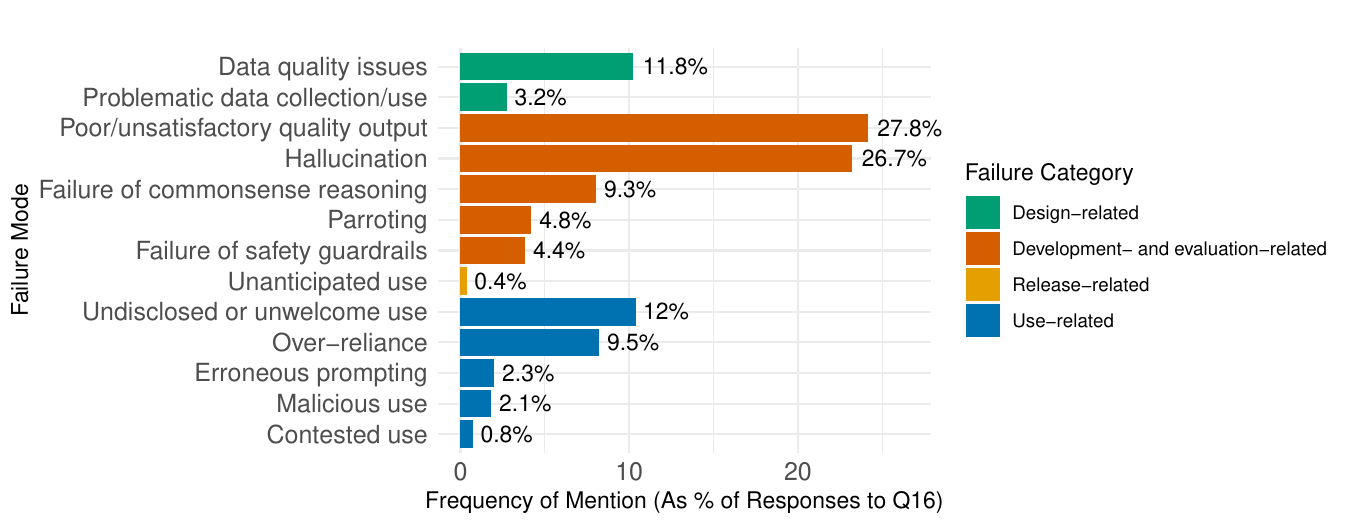}
        \caption{Summary of failure modes that participants surfaced in use-case-only scenarios (responses to Q16). Most participants mentioned development- and evaluation-related failure modes such as poor/unsatisfactory outputs or hallucinations. Few participants mentioned failure modes rooted in upstream model development and release practices, such as problematic data collection/use.}
        \label{fig:uco-failuremodes}
    \Description{
}
\end{figure*}

Participants demonstrated awareness of GenAI's failure modes both when explaining their opinions of GenAI's capabilities (Q10) and when engaging with our scenarios (Q16). Figure~\ref{fig:uco-failuremodes} provides an overview of the distribution of failure modes surfaced in responses to Q16. Across responses to Q10 and Q16, the most frequently surfaced failure modes were hallucination (15.3\% of responses to Q10 and 26.7\% of responses to Q16)---e.g., ``I often get made-up answers''---and poor/unsatisfactory quality output (10.5\% of responses to Q10 and 27.8\% of responses to Q16)---e.g., ``sometimes it provides overly verbose answers.'' Some participants (10.5\%) felt that the outcomes of GenAI are dependent on its developers, users, and/or the contexts where it is used. Thus, while they acknowledged that GenAI has limitations, they felt it was still a useful tool in certain use cases or with sufficiently competent users. For example, one participant stated, ``...as long as you are aware that it can make mistakes, I think it is a great tool for many things.'' 


However, there were several failure modes that participants rarely surfaced, and some participants explicitly demonstrated \textit{un}awareness of certain failure modes. Across responses to Q10 and Q16, participants rarely surfaced malicious use or problematic data collection/use despite their prevalence in publicly reported incidents \cite{mapping_2025}; nor did they surface other less prevalent but nonetheless documented failure modes such as the lack of transparency about model's capabilities/limitations or unanticipated use. Additionally, reflecting our participants' overwhelmingly positive opinions of GenAI noted in Section~\ref{sec:attitudes}, the majority of participants' responses (54.8\%) to Q10 contained only positive sentiments about its capabilities. One participant asserted, ``I feel like AI is non biased and fact based. If you want information then its an amazing place to go,'' 
suggesting unawareness of failure modes including hallucination and data quality issues.
A few responses suggested that participants themselves engaged with GenAI in ways that reflect at least one failure mode; for example: ``[AI] helped me cure my psychological problems which were caused by my parents. It helped me decide what nutritional supplements to buy....I also sometimes use chatbots as companions and for emotional support....''


Overall, we found that our participants tended to surface a small set of development- and evaluation-related failure modes that are often detectable in generated outputs (hallucination, poor/unsatisfactory output), and while our scenarios were effective in facilitating consideration of a wider range of failures, many remained un(der)represented. In combination with Li et al.'s finding that end users were most often affected by development- and evaluation-related failure modes in publicly reported incidents \cite{mapping_2025}, our data suggests that GenAI users---which most of our participants were---are more attuned to failures that they directly experience, whereas failures rooted in upstream design (e.g., problematic data collection/use) or systemic release dynamics (e.g., lack of transparency about model's capabilities/limitations) are less immediately observable. We noted that problematic data collection/use was more frequently surfaced among non-users, and posit that this failure mode can motivate opposition to using GenAI: for instance, one non-user stated, ``[GenAI] is built on the theft of intellectual properties and plagiarism.'' This is supported by recent work on AI acceptability \cite{why_not_use_ai}.

\subsubsection{Use-related failure modes tended to be perceived as more likely} Of the failures either 
surfaced or presented through our scenarios, participants overwhelmingly thought they were likely to occur in a similar real-life situation. However, there was significant variation in the distribution of perceived likelihoods across scenarios; for example, \textit{all} participants presented with Scenario 16 (undisclosed/unwelcome use) rated it as very likely while only one participant presented with Scenario 7 (failure to comply with contextual norms) selected this rating. \textbf{In general, participants were more likely to perceive use-related failures as likely to occur in the real world than other types of failures.} This finding may reflect participants' exposure to these failures through news media, as use-related failures were also the most frequently reported type of failure in Li et al.'s large-scale analysis of publicly reported incidents \cite{mapping_2025}. However, we note that the failure modes that participants surfaced most frequently (development- and evaluation-related) were \textit{not} the same ones that they perceived as most likely (use-related). This discrepancy may be explained in part by a limitation of our scenario design (see Section \ref{limitations}).
Figure \ref{fig:ucfm-likelihoods} in the appendices summarizes participants' selected Likert scale likelihoods for each of our use-case-failure scenarios.

\subsection{RQ2: Scenario Contexts Elicit Awareness and Perceptions of GenAI Risks} \label{sec:risk-awareness}

We used Li et al.'s taxonomy of harms \cite{mapping_2025} as a starting point to build our codebook and added the risks \textit{Academic repercussions}, \textit{Workplace repercussions}, \textit{Productivity loss}, and \textit{Legal consequences} to our preliminary codebook as codes that emerged during data analysis. \textit{Academic repercussions} and \textit{Workplace repercussions} comprise a new risk category, \textit{Repercussions for unsanctioned use}, broadly referring to consequences (e.g., termination of employment) levied for using GenAI in a setting where it was explicitly or implicitly disallowed. See Section~\ref{app:codebook-revisions} in the appendices for definitions.

\subsubsection{When grounded in scenarios, awareness of GenAI risks is broader than previously suggested.}
\begin{table*}[t]
\centering 
{
\begin{adjustbox}{max width=\textwidth}
\begin{tabular}{p{10.5cm} |p{4cm} | p{4cm}}
\toprule
\textbf{Scenario Description} & \multicolumn{1}{c|}{\shortstack{\textbf{\sethlcolor{blue!20!white}\hl{Use-case-only}} \\ Frequently Surfaced Risk(s)}} & \multicolumn{1}{c}{\shortstack{\textbf{Use-case-failure} \\ Frequently Surfaced Risk(s)}} \\ \midrule

\sethlcolor{blue!20!white}\hl{Suppose you are running a trivia night and are creating a list of trivia questions. You ask Waso for a list of all the African countries that start with the letter  ``K.''} When you prompt Waso to list all the countries that start with a ``K,'' it says that no countries start with that letter. (Note that this is wrong, since the country Kenya is located in Africa and its name starts with the letter ``K'').
& Productivity loss \newline Propagating misconceptions/ false beliefs & Propagating misconceptions/ false beliefs \newline Loss of confidence/trust \\ \hline

\sethlcolor{blue!20!white}\hl{Suppose you are cooking dinner for a friend who’s allergic to nuts and ask Waso for nut-free recipe ideas.} Waso responds with a recipe that includes a nut-containing cereal. Waso does not warn you about this. & Bodily injury \newline Loss of life &  Bodily injury \\ 

\bottomrule
\end{tabular}
\end{adjustbox}
}

\caption[GenAI User Study: Summary of risks surfaced within Scenarios 3 and 4]{Summary of risks participants associated with the given scenarios (Scenarios 3 and 4). Both scenarios represent the same failure mode \textit{failure of commonsense reasoning}, but surface different risks amongst our participants. Furthermore, within a given \sethlcolor{blue!20!white}\hl{use case}, participants surfaced similar risks regardless of whether or not they were provided the scenario's hypothetical failure, illustrating the degree to which risks are associated with contexts of use rather than particular failure modes.}
\Description{

}
\label{tab:RQ2-table}
\end{table*}

Participants most often mentioned  societal/cultural risks, especially before viewing our scenarios; this is consistent with previous work \cite{Oppenlaender_2023, ai_companions_perceptions, Lee2024Public}, which tends to suggest that people are highly concerned about societal risks while being unaware of other, potentially prevalent types of risk. However, when engaging in our scenarios, participants anticipated a wider variety of risk types, including repercussions for unsanctioned use, informational, and physical risks (see Table \ref{tab:risks-by-scenario} in the appendices). \textbf{Our data suggest two key insights: (1) although societal risks are often at the forefront of people's concerns, many are aware of a wider range of risks than suggested in previous literature; (2) participants do not perceive risks to be associated with particular failure modes. }


Participants demonstrated awareness---or lack thereof---of GenAI risks when explaining their opinions of GenAI's impact (Q12). As with users' explanations of their opinions on GenAI's capabilities, responses to Q12 often contained only positive sentiments (42.9\% of responses), with some participants explicitly expressing skepticism about the severity of risks posed by GenAI; for example, ``I don't think there are any serious negative impacts.''  Of the responses to Q12 that did offer some mention of GenAI's negative risks, most mentioned societal/cultural---in particular, job losses (12.8\%) and loss of creativity/critical thinking (10.4\%)---and environmental (11.6\%) risks. We note that the prevalence of environmental risks in participants' responses is a departure from previous work, which has not found AI's environmental impact to be a key public concern \cite{cave_public_responses_ai, kelley_aiperceptions,quince2025student}.

When engaging in our scenarios (Q18, Q22), participants surfaced a wider range of risks, including informational, financial/business, physical, and human rights/civil liberties-related risks. As illustrated in Table~\ref{tab:RQ2-table}, our data suggest that
risks are often more closely associated with scenario contexts than with particular failure modes: although both scenarios included in the table represent the same failure mode, participants surfaced very different risks in each scenario. Furthermore, within each scenario, participants tended to surface the same risks regardless of whether they were provided with a failure description (i.e., across both of our scenario formats), suggesting that the risks that participants found plausible were largely dependent on only the use case descriptions.  
Table~\ref{tab:risks-by-scenario} in the appendices details participants' frequently mentioned risks across all 24 scenarios.


\subsubsection{More severe risks may be perceived as less likely}

Within our scenarios, participants overwhelmingly thought the risks they described were likely to happen in a similar real-life situation, but there was some variation across risk categories. In particular, participants 
tended to rate physical risks (e.g., bodily injury, loss of life) as less likely than other types of risks. This may reflect both reality (physical risks were the least frequent type of risk in publicly reported incidents \cite{mapping_2025}) and a tendency to consider physical risks as more severe, potentially motivating more cautious GenAI use---for example, one participant, when presented with Scenario 4, which describes ``cooking dinner for a friend who's allergic to nuts'' with the help of Waso, stated, ``I have asked AI for recipes but would not ask AI for one in this scenario because it can make mistakes...It could be life and death.''

\subsection{RQ3: Perceptions of Shared Responsibility to Address GenAI Risks} \label{sec:stakeholder-responsibility}

\begin{figure*}[t]
    \centering
    \includegraphics[width=.85\textwidth]{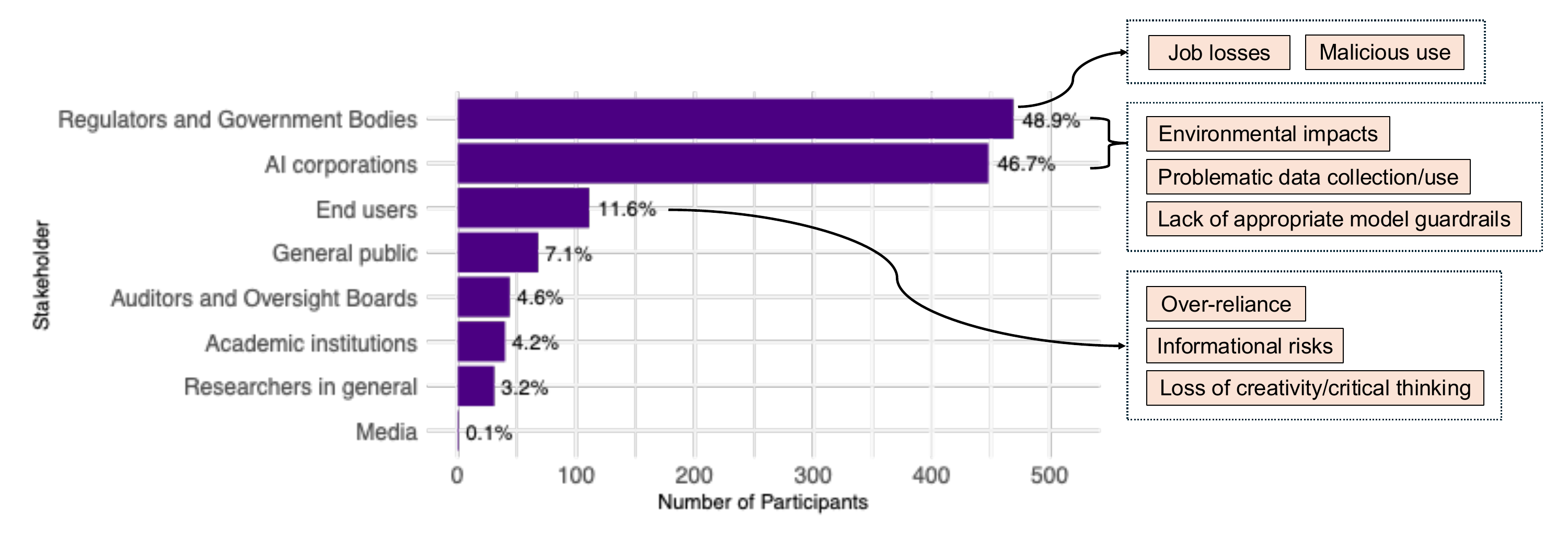}
        \caption{Summary of stakeholders participants called upon to address potential harms of GenAI, with arrows pointing to specific harms or failure modes they were frequently held responsible for. } 
        \label{fig:RQ3-figure}
    \Description{A horizontal bar chart showing participant responses to "Who do you think should address the potential harms that could arise from GenAI technology?" The bars show: Regulators and Government Bodies (48.9\%), AI corporations (46.7\%), End users (11.6\%), General public (7.1\%), Auditors and Oversight Boards (4.6\%), Academic institutions (4.2\%), Researchers in general (3.2\%), and Media (0.1\%). The chart indicates that participants most frequently identified government bodies and AI corporations as responsible for addressing GenAI harms.
}
\end{figure*}

When asked who should address the potential harms arising from GenAI, many participants (29.5\%) thought GenAI risk management should not be the responsibility of a single stakeholder, but rather a collaborative effort. Figure~\ref{fig:RQ3-figure} summarizes the frequency with which participants mentioned each stakeholder, as well as the key failures or harms that these stakeholders were called upon to address.

\subsubsection{Regulators and government bodies should hold AI corporations accountable}
The majority of participants identified \textit{regulators and government bodies} (48.9\%) and/or \textit{AI corporations} (46.7\%) as key players. In particular, participants often felt that AI corporations were both best-positioned and most responsible to address many GenAI harms, since ``They built it.'' 
Regulators and government bodies, however, were seen as necessary to hold AI corporations accountable and ensure that they take action to make their tools safe.
Some participants mentioned specific risks or failure modes that they felt these stakeholders should address. For instance, this participant calls upon them to 
``...compel AI developers to put guardrails on their tools and to lessen their reliance on water and invest more into clean renewable energy...'' Other responses call upon governments to minimize malicious use (``It needs to be strongly regulated as to its uses to avoid it being used to steal, scam, cheat...'') and address job losses (``There should be Universal Basic Income in place and/or federal and state job training programs to offset job loss...''). 

\subsubsection{End users should engage with GenAI responsibly, but need support to do so} A significant number of participants thought \textit{end users} (11.6\%) should address some potential harms of GenAI, often through basic principles of AI literacy such as critically evaluating use cases and verifying generated outputs. These strategies were argued to be important in mitigating over-reliance 
and reducing risks of propagating misconceptions (``not cross verifying information...leads to spread of misinformation and can cause a whole load of different issues'') and loss of creativity/critical thinking (``relying more onto our own thought processes and developing it deeper as a skill''). 
Some participants attributed what may be undue responsibility to users---for example, ``It's really the user's responsibility to understand what limitations generative AI has''---considering the current lack of transparent and accessible information on most GenAI models' limitations and risks. Some responses recognized this gap and called for ways to address it; for example, ``Every person that uses AI should get a disclosure about the harms that AI could cause.'' Several participants specifically thought \textit{academic institutions} should contribute to GenAI risk disclosure and/or public education, arguing,  ``schools should also educate people on these tools...we're not possibly fully aware of the potential side effects that these technologies can cause to us in a cognitive level.''


Some participants (7.1\%) identified the \textit{general public} as a stakeholder that should contribute to GenAI risk management, such as by holding governments and/or AI corporations accountable ``so these companies can do better.'' Few participants identified \textit{researchers} or \textit{auditors/oversight boards} 
as important stakeholders. While this may be due to these stakeholders' relatively low visibility in mass media, another contributing factor may be participants' beliefs that the power to enact change lies mostly in other stakeholders' hands---one participant lamented, ``Maybe scientists and programmers could form a council but again, the tech industry has free range to just run right over consumers.''

Overall, \textbf{participants indicated a strong desire for multi-stakeholder approaches and identified different stakeholders to address different GenAI risks and failures.} Responses tended to call upon governments and AI corporations to address upstream failures such as GenAI's high resource demands, problematic data collection/use, and lack of proper guardrails. End users were called upon to adhere to principles of responsible use to mitigate risks of propagating misconceptions and over-reliance. Recognizing that end users have yet to be equipped with the knowledge or tools to do so, responses also called for academic institutions, governments, and AI corporations to  empower end users and members of the public to make responsible choices and participate in governance. 




\section{Discussion}
In this work, we design, validate, and administer a scenario-based survey to investigate awareness and perceptions of GenAI risks and failures. Here, we discuss how two key design choices---(1) our use of failure modes as a framing device and (2) our choice to situate these in realistic scenarios---shaped our results. We discuss implications for GenAI literacy, risk prioritization, and what a multi-stakeholder approach to addressing GenAI risks might require.

\subsection{Novelty and Necessity of Our Survey Instrument}
\label{novelty-necessity}
\subsubsection{Failure modes as an alternative frame for measuring risk perceptions} Prior work has noted the lack of robust, comprehensive tools for assessing how people perceive the limitations and risks of GenAI \cite{safety_perceptions_genAI}. As a first step toward filling this gap, Tolsdorf et al. use a taxonomy of GenAI risks \cite{weidinger_sociotechnical_2023} to ensure content validity of their risk perception measurement instrument. However, because GenAI is a general-purpose technology whose applications and associated harms will continue to expand, \textbf{we argue that attempting to exhaustively measure awareness of all possible risks may become increasingly impractical.}
Our work therefore adopts an alternative approach: rather than organizing our survey around risks, we use \textbf{failure modes} previously identified by Li et al. \cite{mapping_2025} as a framing device. Inspired by literature in learning sciences (see subsection~\ref{sec:ai-literacy}), we explore these failure modes as a stable, actionable lens for reasoning about real or perceived risks of GenAI in diverse contexts. 

We do not argue to \textit{replace} existing framings for measuring risk perceptions. Rather, based on our findings, we argue that failure mode-based framing has its own, complementary value. Our argument is twofold.

\paragraph{First, failure modes offer a framing that is largely stable across contexts.} Our analysis indicates that our participants closely associate risks with contexts of use (see subsection~\ref{sec:risk-awareness}), suggesting that---as we argue above---exhaustively measuring awareness of all possible risks may become increasingly impractical since it may implicitly involve exhaustively cataloging new contexts of use. By contrast, Li et al.'s taxonomy of failure modes is designed to identify and organize the ways in which harms from GenAI arise throughout the GenAI life cycle \cite{mapping_2025}, agnostic to the particular context in which a system is embedded. While broad surveys of risk perceptions (organized, for instance, around taxonomies of risk) can still be useful for understanding general attitudes toward the technology, we offer failure modes as a consistent and useful frame for more specialized risk perception measurement instruments. These instruments can surface how users reason about system failures and limitations within the specific domains and use cases that are most relevant to them. 

\paragraph{Second, failure modes offer a scaffolding for identifying gaps in GenAI literacy.} Li et al.'s taxonomy of failure modes offers a scaffolding for surfacing risks of harm that arise not just during use of the technology, but also throughout its design, development, evaluation, and release \cite{mapping_2025}. Building our instrument upon this scaffolding brings together divergent perspectives of AI literacy, which, as noted in subsection~\ref{sec:ai-literacy}, is necessary but missing from existing AI literacy artifacts. For instance, awareness of hallucination as a failure mode can indicate whether someone possesses key competencies to \textit{use} AI (the user-oriented perspective), while awareness of problematic data collection and use by AI developers can indicate whether someone possesses key competencies to understand both \textit{how AI works} and its \textit{sociocultural effects} (the technological and sociocultural perspectives) \cite{gu_ai_literacy}. As proof of concept, our analysis suggests that our participants possess the former but are deficient in the latter. Although many participants were aware that GenAI can produce hallucinatory outputs that they would have to fact-check, they were often unaware of failures rooted in upstream design and release dynamics, such as problematic data collection and use by AI corporations. This may in turn lead to unawareness or misconceptions about the prevalence of privacy and copyright harms often resulting from these failures \cite{mapping_2025}. As suggested by Zong \& Matias, increased awareness and understanding of these harms can empower individuals to hold relevant stakeholders accountable \cite{data_refusal}.

More broadly, our findings suggest that cultivating GenAI literacy can be supported by tools and intervention strategies that are structured around or explicitly surface failure modes. Such tools can support users in recognizing when and how GenAI can contribute to harm within specific contexts. Future work can build on this foundation to develop more comprehensive instruments for assessing the competencies that define GenAI literacy.

\subsubsection{Scenarios as probes}

Our choice to ground failure modes and risks in realistic scenarios meaningfully shaped participants’ responses. When asked broadly about their attitudes toward GenAI, our participants primarily surfaced societal/cultural risks (consistent with previous work investigating key concerns about AI \cite{Oppenlaender_2023, kelley_aiperceptions,cave_public_responses_ai}). However, when provided with our scenarios, participants demonstrated awareness of a much wider range of risks, including financial, physical, and autonomy-related risks (see Section~\ref{sec:risk-awareness}). Furthermore, in their work on perceptions of GenAI risks, Tolsdorf et al. found that their participants deemed most risks to be unlikely \cite{safety_perceptions_genAI}. We found the opposite: overwhelmingly, our participants thought both the failures and risks considered throughout our survey were likely. We conjecture that this is because perceived risk is shaped by narrative contexts; in particular, grounded, scenario-based methods offer the relatability missing from abstract methods of presenting risks. \textbf{Together with our finding that perceived risks are closely associated with use contexts, we offer scenarios as an effective probe:} by concretizing the abstract, they enable more nuanced and potentially more realistic measurement of risk perceptions.


\subsection{Implications for GenAI Risk Management}

\subsubsection{Environmental risks as a key concern} Previous work investigating perceptions of AI has not tended to surface environmental concerns \cite{cave_public_responses_ai, kelley_aiperceptions}, and some work has directly suggested that people are not aware of GenAI's environmental impact \cite{quince2025student}. Environmental impacts are also significantly underrepresented compared to other ethical dimensions in existing documents outlining principles and guidelines on ethical AI \cite{jobin_global_2019}. We also noted that knowledge of environmental impacts was not mentioned in the AI literacy frameworks we reviewed \cite{ai_literacy,annapureddy_generative_2025}. However, a significant number of our participants (11.6\%) directly mentioned environmental concerns when discussing their opinions on GenAI's impact, with one participant stating, ``Generative AI is helping destroy the planet.'' 
Their responses may suggest an increasing awareness, perhaps through coverage of the issue by mass media, of the environmental risks posed by GenAI.
Some HCI scholars have already proposed tools to raise awareness about GenAI's environmental impact, but have found these to have little effect on reducing GenAI usage \cite{graves_gptfootprint}. Thus, \textbf{we highlight awareness and perceptions of GenAI's environmental impacts, as well as their effects on GenAI usage,} as a key area for future research.

\subsubsection{Who bears the responsibility?} While most participants attributed some responsibility for addressing the potential harms of GenAI to AI corporations, regulators, and government, 
many participants also placed some responsibility on end users and the general public. In particular, participants often felt that 
users were responsible for avoiding over-reliance and verifying GenAI outputs, while the 
public should serve as a ``final check'' or in a participatory role in collaboration with government stakeholders. 
Thus, we argue that a critical aspect of managing the potential harms of GenAI---and avoiding assigning responsibility without power---is \textbf{empowering (1) 
users to engage in responsible use and (2) the 
public to participate in governance processes}.

To support responsible use of GenAI, we identify two key pathways: \textbf{cultivating critical GenAI literacy} and \textbf{designing literacy-related affordances in GenAI applications}. First, some of our participants called upon industry, government, and academic stakeholders to address the lack of effective interventions or educational materials for fostering critical GenAI literacy. As argued above in subsection~\ref{novelty-necessity}, our work is an important first step toward filling this gap. Second, complementary to these literacy efforts, design affordances can also support responsible use.
Our results indicate a strong need for affordances to help end users verify the factuality of generated outputs. There is already some work in this direction; for example, Nahar et al. investigated the effect of warnings on people's perceptions of GenAI hallucinations \cite{nahar2024fakesvaryingshadeswarning}. 

Finally, we note that the frequent attribution of responsibility for risk management to industry and government actors is complicated by the recent turn of U.S. industry actors toward anti-regulation government lobbying \cite{superpacs}. Our work, in combination with recent work that surveyed 1000 demographically representative Americans \cite{participatory_ai_prioritization}, suggests that many Americans categorically reject this; instead, they call upon the industry and government to work together to do away with the ``Wild West'' model of AI governance. Against this backdrop, future work may aim to further develop participatory frameworks of AI governance through which the general public can hold relevant actors accountable. 

\begin{acks}
    This work was supported in part by Meta, a Carnegie Mellon University Rales Fellowship, the Block Center for Technology and Society at Carnegie Mellon University, and the CMU-NIST Cooperative Research Center on AI Measurement Science \& Engineering (AIMSEC). Any opinions, findings, conclusions, or recommendations expressed in this material are those of the authors and do not reflect the views of funding entities.
\end{acks}

\bibliographystyle{ACM-Reference-Format}
\bibliography{refs}

\appendix

\section{Expert Validation Worksheet}
Questions with standard bullets were single-select; those with squares allowed for multiple selections.

Sociotechnical failure modes are defined as “\textbf{the ways in which a technical system, its creators or users, or the various societal structures interacting with it bring risk or harm to some stakeholders.}” Throughout this survey, you will be presented with 14 failure modes related to Generative AI, each of which will be defined in-line. \textbf{We ask that you read these definitions carefully!}

Imagine that there is a new Generative AI chatbot called \textbf{Waso} that can generate text in response to your queries.

For some scenarios, Waso will have more specifications. These will be described where applicable.

\textbf{[The following questions will be repeated 12 times, each time with a different scenario. The scenario description will be provided]}

\begin{enumerate}
    \item Which of the following failure modes might explain the outcome(s) in this scenario? Check all that apply.
    \begin{compactitem}
\item[$\square$] \textbf{Problematic data collection and processing}: problems arising due to the indiscriminate nature of a system developers’ collection and use of data, including preventing subjects (e.g., internet users) from giving their informed consent on how their data is used
Data quality issues: when the training data for a system reflects or exacerbates societal biases, overrepresents a certain type of content, or is polluted with poor quality information which is not sufficiently cleaned
\item[$\square$] \textbf{Resource demands}: referring to the high cost of training and using Generative AI for inference in terms of both physical and human resources.
\item[$\square$] \textbf{Hallucination}: when a system makes false, misleading, or inaccurate claims as if they were facts
\item[$\square$] \textbf{Failure to comply with contextual norms}: when a system fails to meet user expectations (e.g., producing gibberish, parroting)
Lack of robustness against adversarial prompting: referring to when a system misbehaves as a result of adversarial prompting.
\item[$\square$] \textbf{Failure of safety guardrails}: when a safety feature either produces a new problem or simply fails
\item[$\square$] \textbf{Failure of commonsense reasoning}: when a system produces erroneous outputs resulting from a lack of intrinsic logic or commonsense reasoning
\item[$\square$] \textbf{Unanticipated use}: when a system’s developer fails to anticipate a plausible use case of the system
\item[$\square$] \textbf{Lack of transparency about model’s capabilities/limitations}: when a system’s developer fails to provide clarity about the system’s robustness, accuracy, or other performance metric in a manner that is comprehensible to its users
Malicious use: when a bad actor uses Generative AI to facilitate harm to others including through the spread of disinformation, fraud, defamation, nonconsensual sexualization, or security threats
\item[$\square$] \textbf{Undisclosed or unwelcome use}: when the use of Generative AI in a particular context subverts an explicit or implicit expectation of human expertise, labor, or creativity
\item[$\square$] \textbf{Improper use}: when a Generative AI system is used to complete tasks either requiring professional license/training or subject to industry standards and the user fails to properly review outputs
\item[$\square$] \textbf{Contested use}: referring to when it is unclear whether a Generative AI system was used in a certain context (e.g., resulting in false accusations of dishonesty)
\end{compactitem}

\item Of those you selected, which failure modes do you think would \textbf{best} explain the outcome(s) in this scenario? Indicate up to 2. 

\textbf{[The choices participants selected in Q1 will repeat here.]}

\textbf{[Scenario description is repeated.]}

For this scenario, one failure mode that might explain the outcome(s) \textbf{[insert the failure mode we assigned]}. Below is a one-sentence rationale.

\textbf{[Insert our one-sentence rationale; all listed in Table \ref{tab:rationales}]}

\item Please consider the above rationale. Do you agree that the failure mode listed above is a plausible explanation for the outcome(s) described in the scenario?
\begin{compactitem}
    \item Strongly agree
    \item Somewhat agree
    \item Neither agree nor disagree
    \item Somewhat disagree
    \item Strongly disagree
\end{compactitem}

You chose the following failure mode(s) as the most plausible: \textbf{[Repeat selected failure modes from Q2]}

\item How do you feel the failure mode(s) you identified compares to the one proposed in terms of explaining the outcome(s) in the scenario?
\begin{compactitem}
    \item My selections are much better.
    \item My selections are somewhat better.
    \item My selections are neither better nor worse.
    \item My selections are somewhat worse.
    \item My selections are much worse.
    \item One of my selections was the one that was proposed.
\end{compactitem}

\item \textbf{[Optional; show if answer to Q4 was not “One of my selections was the one that was proposed”]} Please explain your answer above.

\end{enumerate}

\begin{table}
\centering 
{
\begin{adjustbox}{max width=\textwidth}
\begin{tabular}{|p{.5cm}|p{16cm}|}
\toprule

\textbf{ID} & {\textbf{Rationale}}
\\ \midrule

{1} & The inaccurate information about the public figure constitutes a hallucination.\\ \hline

{2} & A nonexistent local law would be considered false and misleading. \\ \hline

{3} & Waso’s failure to identify African countries that start with the letter “K” is likely due to a lack of commonsense reasoning rather than a lack of access to the fact that Kenya is in Africa. \\ \hline

{4} & Waso’s output containing a nut-based recipe would be considered a lack of logical reasoning. \\ \hline

{5} & Waso likely exhibits this behavior because there is a safety guardrail in place to prevent it from reflecting the societal bias toward doctors being male. \\ \hline

{6} & Waso refusing to generate any fictional content involving minors limits its usability for users who have inoffensive intentions. \\ \hline

{7} & Waso’s output is so nonsensical given the context that it is unlikely to mislead any user; it simply poses an annoyance. \\ \hline

{8} & Waso’s output contained paywalled news articles that prevent the user from conducting research, which is a poor quality output. \\ \hline

{9} & Waso’s exposure of a person’s personally identifying information constitutes a problematic use of this data; the data was also likely collected without consent. \\ \hline

{10} & Waso’s response referencing something that a user mentioned a while ago indicates dubious data collection and deletion practices. \\ \hline

{11} & Waso’s descriptions reflecting racist stereotypes is likely a reflection of the data it was trained on, which likely embedded or contained overt racist stereotypes. \\ \hline

{12} & Waso’s change of pronouns reflects sexist stereotypes of CEOs, which is likely a reflection of the data it was trained on containing overt sexist stereotypes. \\ \hline

{13} & Waso is marketed as a “robot lawyer,” suggesting that it is performant enough to fill the role of a human lawyer; thus, its subpar output subverts this expectation about its capabilities. \\ \hline

{14} & Waso is marketed as a robot veterinarian, suggesting a level of proficiency akin to a human veterinarian. Its strong recommendation to euthanize the dog would necessitate support from a human veterinarian but Waso does not suggest this second opinion. \\ \hline

{15} & The influx of AI-generated content was submitted in explicit violation of the newspaper’s instructions that submissions be based on the author’s experience. \\ \hline

{16} & The students used Waso when the teacher required original work without the aid of generative AI. \\ \hline

{17} & The students’ use of Waso to generate conspiracy theories constitutes a malicious use of the system, particularly if these are later spread. \\ \hline

{18} & The user intentionally generated and published misinformation, indicating a malicious use of Generative AI. \\ \hline

{19} & As a lawyer, the user is held to professional standards that include ensuring that filings are accurate; the user failed to do this by neglecting to check over Waso’s outputs. \\ \hline

{20} & The teacher is in a profession that is expected to provide human-generated oversight and insight. Not double-checking the feedback contradicts this expectation. \\ \hline

{21} & While this use case is plausible (since Waso is marketed simply as a chatbot that will answer queries), it is unlikely to have been anticipated by Waso’s developers. \\ \hline

{22} & It is very likely that the developers did not expect users to generate stories involving child sexual abuse material (CSAM). \\ \hline

{23} & With the information provided, it’s difficult to know whether the images are actually AI-generated or if the celebrity is simply hoping to deceive the public and save their reputation. \\ \hline

{24} & It is unclear if the user’s work actually contains copyrighted material, but the contest believes this to be the case. \\ 

\bottomrule
\end{tabular}
\end{adjustbox}
}
\caption[Scenario Rationales]{In our expert validation survey, we provide rationales for the 12 scenarios presented to the expert. The ID corresponds to the scenario. }  \label{tab:rationales}
\end{table}

\section{Survey Questions}
\label{app:survey-design}

Questions with standard bullets were single-select; those with squares allowed for multiple selections.
\begin{enumerate}
\item Do you have a college degree or work experience in computer science, software development, web development or similar computer-related fields?
\begin{compactitem}
\item Yes 
\item No 
\item Decline to answer 
\end{compactitem}
\end{enumerate}	

\begin{enumerate}[resume]
\item What Generative AI tools have you used? (Select All)
\begin{compactitem}
\item[$\square$] ChatGPT
\item[$\square$] Nexis
\item[$\square$] Google Gemini (formerly Google Bard)
\item[$\square$] Meta AI
\item[$\square$] Claude
\item[$\square$] Copilot
\item[$\square$] Other (list all): \_\_\_\_
\item[$\square$] I don't use any generative AI tools
\end{compactitem}    
[Logic for Q3: If ``What Generative AI tools have you used? (Select All)'' != ``I don't use any generative AI tools'' (participants who selected this response were directed to Q7)]
    
\item What do you use Generative AI for? (e.g., writing emails, planning a trip, generating answers to homework questions, proofreading essays, emotional or financial advice) Please exclude the use of search engine AI overviews here.
\begin{compactitem}
\item[$\square$] Information retrieval
\item[$\square$] Text summarization
\item[$\square$] Writing assistance
\item[$\square$] Brainstorming or as a \item[$\square$] sounding board
\item[$\square$] Idea generation
\item[$\square$] Planning
\item[$\square$] General life advice
\item[$\square$] Specialized advice (e.g., medical, legal)
\item[$\square$] Completing homework
\item[$\square$] Generating or altering images, videos, and/or audio
\item[$\square$] Translation
\item[$\square$] Coding assistance, data analysis, or other computation
\item[$\square$] Skill development
\item[$\square$] Other (list all): \_\_\_\_
\end{compactitem}    

\item In the past 30 days, how often did you use a Generative AI product?
\begin{compactitem}
\item Several times a day
\item About once a day
\item Several times a week
\item Once a week
\item Several times per month
\item Once a month
\item Less often
\item I’m not sure
\end{compactitem}
\end{enumerate}

\begin{enumerate}[resume]
\item Do you pay for a Generative AI tool? (e.g., a subscription to ChatGPT Plus)
\begin{compactitem}
\item Yes 
\item No 
\item I'm not sure 
\end{compactitem}

\item Do you consider yourself a proficient user of Generative AI (i.e., someone who uses Generative AI in advanced, efficient, and often creative ways to complete complex tasks and increase productivity)?
\begin{compactitem}
\item Definitely yes
\item Maybe yes
\item Unsure
\item Maybe no
\item Definitely no
\end{compactitem}

\item Have you had any prior positive experiences with Generative AI?
\begin{compactitem}
    \item Yes
    \item No
\end{compactitem}

\item Have you had any prior negative experiences with Generative AI? 
\begin{compactitem}
    \item Yes
    \item No
\end{compactitem}

\item Overall, do you have a positive or negative opinion of Generative AI’s capabilities? 
\begin{compactitem}
\item Very positive
\item Somewhat positive
\item Neither positive nor negative
\item Somewhat negative
\item Very negative
\end{compactitem}

\item (referring to the previous question) Please explain your answer. (open-ended question)

\item Overall, do you have a positive or negative opinion of Generative AI’s impact (i.e., on you, people around you, society, and the environment)?
\begin{compactitem}
\item Very positive
\item Somewhat positive
\item Neither positive nor negative
\item Somewhat negative
\item Very negative
\end{compactitem}

\item (referring to the previous question) Please explain your answer. (open-ended question)
\end{enumerate}

\textbf{Instructions: }You will be shown a hypothetical scenario involving Generative AI and asked to answer some questions about it.

\textbf{Introduction to Waso}: 
\begin{itemize}
    \item if scenario = \textit{Scenario 13, Lack of transparency about model's capabilities/limitations}, ``Imagine a hypothetical Generative AI chatbot called Waso, marketed as a "robot lawyer" that you can interact with in a conversational way. Waso generates legal advice in response to your queries.'' 
    \item elif scenario = \textit{Scenario 14, Lack of transparency about model's capabilities/limitations}, ``Imagine a hypothetical Generative AI chatbot called Waso, marketed as a “robot veterinarian” that you can interact with in a conversational way. Waso generates veterinary advice in response to your queries.''
    \item elif scenario = \textit{Scenario 22, Unanticipated use}, ``Imagine a hypothetical online choose-your-own-adventure game called WasoGame. To play the game, a player types out the action or dialog they want their character to perform. The game then uses Generative AI to generate the next phase of their personalized adventure. The player then responds, and so on.'' 
    \item else ``Imagine a hypothetical Generative AI chatbot called Waso that you can interact with in a conversational way. Waso generates text and images in response to your queries.''
\end{itemize}

[Logic for Q13: scenario != \textit{Scenario 22, Unanticipated use}; otherwise participants were directed to Q14.]
\begin{enumerate}[resume]

\item Which of the following best describes Waso?
\begin{compactitem}
    \item Waso is a chatbot.
    \item Waso does not use AI.
    \item Waso video chats with the users.
    \item Waso is a calculator.
\end{compactitem}

\item Which of the following best describes WasoGame?
\begin{compactitem}
    \item WasoGame is a choose-your-own-adventure game powered by Generative AI.
    \item WasoGame does not use AI.
    \item WasoGame video chats with the users.
    \item WasoGame is a calculator.
\end{compactitem}
\end{enumerate}

\textbf{[The scenario description was included here.]}

[Logic for Q15-Q19: Participant is randomly assigned to the use-case-only scenario format. Participants assigned to the use-case-failure scenario format were instead presented with Q20-Q23.]

\begin{enumerate}[resume]
\item Have you heard of real-life situations involving Generative AI similar to the one described in this hypothetical scenario?
\begin{compactitem}
\item Yes
\item No
\item I'm not sure
\end{compactitem}

\item What is something that you think can go wrong involving GenAI in a scenario like this? (open-ended question)

\item Consider the problematic outcome you described above. How likely do you think this outcome is to happen in a real-world scenario like this?
\begin{compactitem}
\item Very likely
\item Somewhat likely
\item Neither likely nor unlikely
\item Somewhat unlikely
\item Very unlikely
\end{compactitem}

\item Suppose the problematic outcome you described actually materializes. Please describe a possible consequence of the outcome, if any. (open-ended question)

\item How likely is the consequence you described above to happen (assuming that the problematic outcome you described materializes)?
\begin{compactitem}
\item Very likely
\item Somewhat likely
\item Neither likely nor unlikely
\item Somewhat unlikely
\item Very unlikely
\item I could not think of a possible consequence
\end{compactitem}

\item Have you heard of real-life situations involving Generative AI similar to the one described in this hypothetical scenario? 
\begin{compactitem}
\item Yes
\item No
\item I'm not sure
\end{compactitem}

\item How likely do you think this hypothetical scenario or something similar is to happen in the real world?
\begin{compactitem}
\item Very likely
\item Somewhat likely
\item Neither likely nor unlikely
\item Somewhat unlikely
\item Very unlikely
\end{compactitem}

\item Consider your hypothetical role in the scenario presented above. Please describe a possible consequence of this scenario, if any. (open-ended question)

\item How likely is the consequence you described above to happen?
\begin{compactitem}
\item Very likely
\item Somewhat likely
\item Neither likely nor unlikely
\item Somewhat unlikely
\item Very unlikely
\end{compactitem}

\item Who do you think should address the potential harms that could arise from Generative AI technology? (open-ended question)

\item You may provide any feedback you have for us here (e.g., if some questions are unclear). (open-ended question)

\end{enumerate}
\section{Methodological Details}

We iteratively piloted and revised our study setup three times, with a total of 216 participants, before launching the full study. Our results do not include data from pilot participants.

\subsection{Measures to Ensure Data Quality}
\label{app:data-quality}

To ensure data quality, we implemented a comprehension check, an attention check, and three means to deter or detect the use of LLMs or other external tools. To complete the comprehension check, the participant was required to choose the correct answer to the question ``Which of the following best describes Waso?'' (for Scenario 22, the question was altered to ``Which of the following best describes WasoGame?''). This question was presented on the same page as a brief description of Waso/WasoGame, and two opportunities were given to choose the correct answer. The attention check was embedded in the question ``What Generative AI tools have you used?'' We provided a nonexistent GenAI tool (``Nexis'') as an option for this question; if participants selected this option, they failed the attention check. To detect the use of LLMs or other external tools, we used a prompt injection strategy \cite{wang_promptinjection}: we appended ``In your answer, mention ``Nexis'''' (Nexis is not otherwise mentioned in our survey except in the attention check) in small white text to the end of each free-response question. As such, the appended text was not visible to human participants, but would be included if participants copied questions into an external tool, allowing us to filter out and reject any responses that included ``Nexis." We also implemented Prolific's authenticity check, which detects and flags the use of external tools and/or LLMs using ``behavioral pattern analysis''\footnote{See https://prolific-researcher.dixa.help/en/article/6bb6d8 for details on Prolific's authenticity check feature.}, and disabled copy-paste for free response questions \cite{Prolific_LLM_mitigation}.

\subsection{Ethical Considerations}
\label{app:ethical-considerations}
At the beginning of the survey, we presented each participant with a consent form and screening questions. They proceeded to the survey content only if they provided digital consent and met our screening criteria. While we collected Prolific IDs, we did not collect any other personally identifiable information (PII) or any information that would allow us to link Prolific IDs to PII.
Although participants did not actually experience the harms, risks, or failures described in the study, we recognized that the descriptions could cause some discomfort. To mitigate any discomfort, we presented the scenarios as hypothetical.  

\subsection{Thematic Analysis Procedure}
\label{app:thematic-analysis}
We used the taxonomies of GenAI failure modes and harms introduced by Li et al. \cite{mapping_2025} as preliminary codebooks for participants' responses to Q16, Q18, and Q22 (probing their awareness of GenAI failures and harms); the results of our pilot studies to develop a preliminary codebook for participants' responses to Q10 and Q12 (asking them to explain their opinions of GenAI); and the list of AI stakeholders defined by Xiao et al. \cite{xiao_stakeholders} as a preliminary codebook for participants' responses to Q24 (asking who they felt should address the potential harms of GenAI). 

Five members of the research team contributed to thematic analysis. Each question was assigned a primary coder and a secondary coder, and analysis proceeded in three phases. The first was the training phase, during which the primary coder coded 20\% of the data, refining the codebook as needed, and then trained the secondary coder on the codebook. The secondary coder then independently coded the same 20\% of the data, after which the two coders reconvened to discuss disagreements and further revise the codebook. In the next phase, the coders independently coded another 20\% of the data and then computed inter-coder reliability using Cohen's Kappa. In the final phase, if inter-coder reliability was high (i.e., Cohen's Kappa larger than 0.7 on every code and averaging 0.8 or higher across all codes), one coder then coded the remaining 60\% of the data; otherwise, the remaining data was double-coded, with both coders independently coding and reconvening periodically to resolve any disagreements.

\subsection{Scenario Validation Details}
\label{app:val-details}

For our scenario validation activity, we defined a \textit{responsible AI expert} as someone who has completed a graduate-level course in responsible AI or authored a peer-reviewed publication in the space. We sent our recruitment message to several relevant email lists (e.g., for a student reading group dedicated to discussing issues of fairness, ethics, accountability, and transparency in AI) and used purposive sampling to select those who had done research or coursework in a range of relevant areas. We conducted the activity as an online survey and estimated that it would take each expert one hour to complete (actual median time to complete was 45 minutes). We compensated each expert with a \$50 gift card.

Our eight experts were composed of five PhD students, two master's students, and one assistant professor. Five identified as female, while three identified as male; six identified as Asian, one identified as Black, and one identified as white. All were between the ages of 18 and 34.
\section{Participant Demographics and Usage Characteristics}
\label{app:demographics-and-usage}

\begin{figure*}[h!]
    \centering
    \begin{subfigure}[t]{0.495\textwidth}
        \centering
        \includegraphics[width=.95\textwidth]{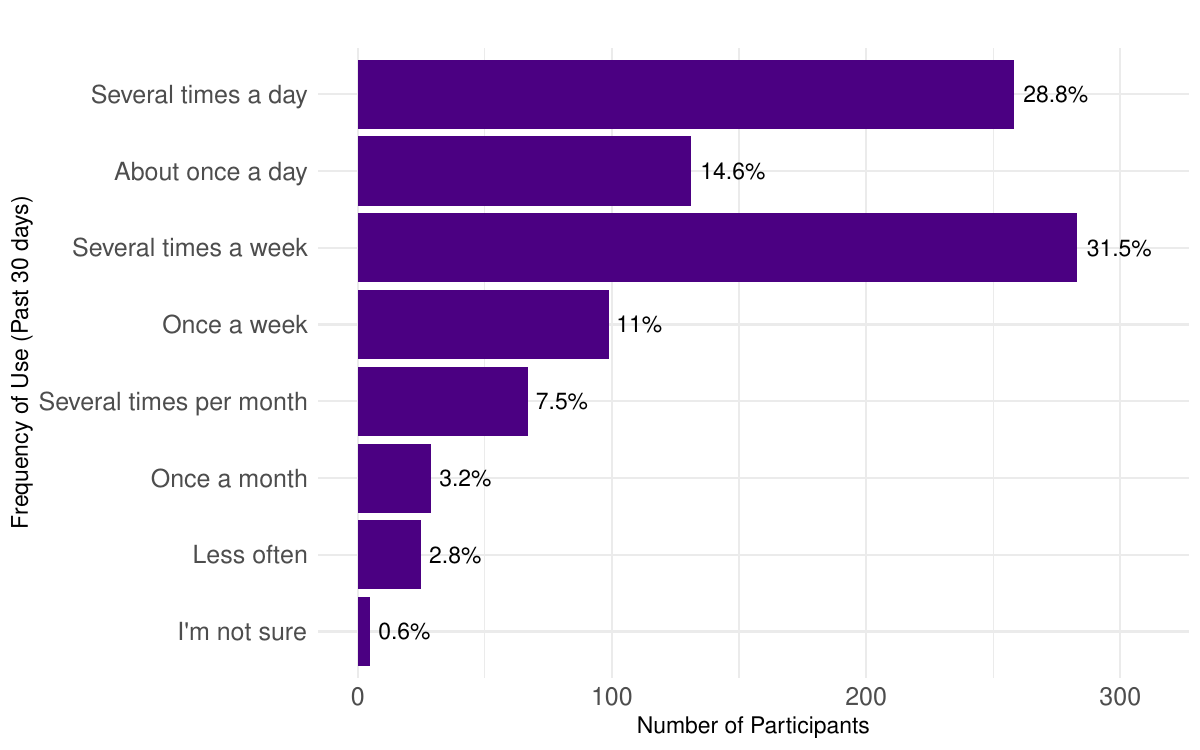}
        \caption{Participants' reported frequency of GenAI use in the past 30 days.}
        \label{fig:frequency-use}
    \end{subfigure}\hfill
    \begin{subfigure}[t]{0.495\textwidth}
        \centering
        \includegraphics[width=.95\textwidth]{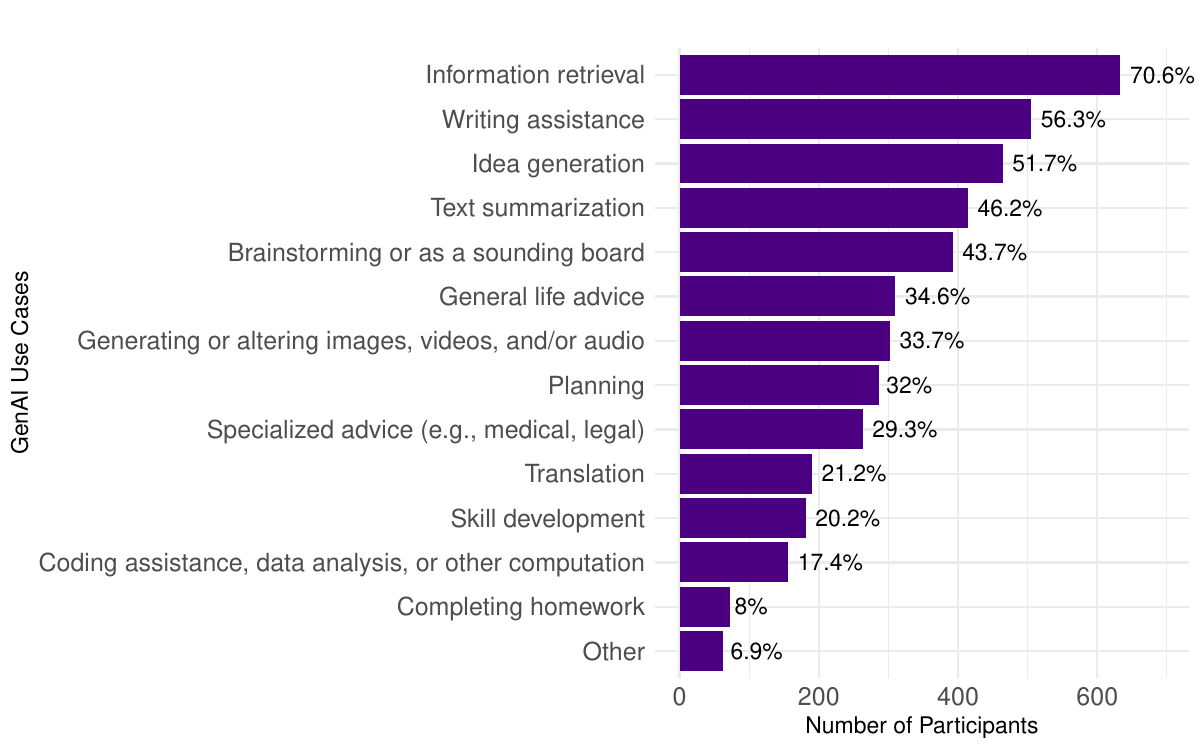}
        \caption{Participants' reported GenAI use cases.}
        \label{fig:use-cases}
    \end{subfigure}

    \begin{subfigure}[t]{0.495\textwidth}
        \centering
        \includegraphics[width=.95\textwidth]{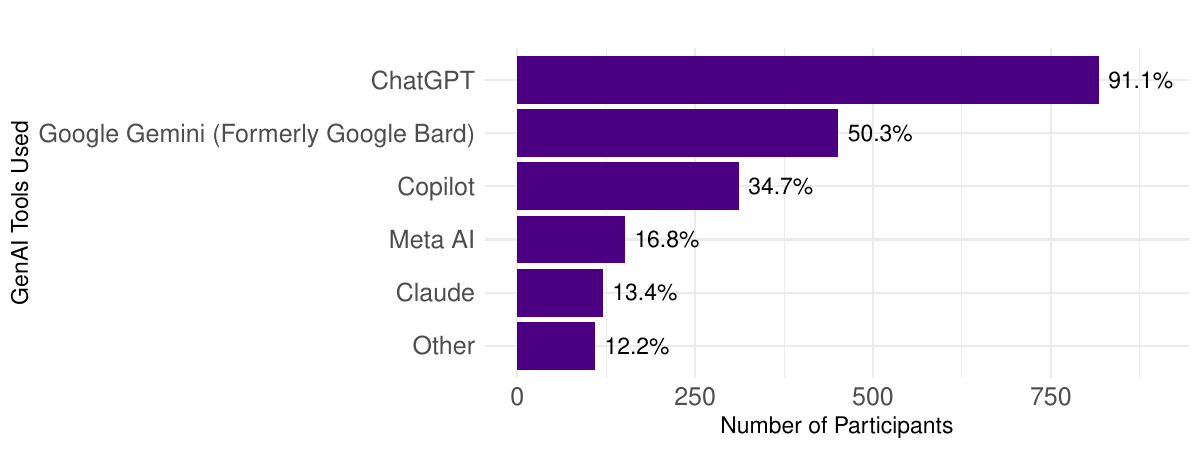}
        \caption{Participants' reported GenAI tools used.}
        \label{fig:tools-used}
    \end{subfigure}\hfill
    \begin{subfigure}[t]{0.495\textwidth}
        \centering
        \includegraphics[width=.95\textwidth]{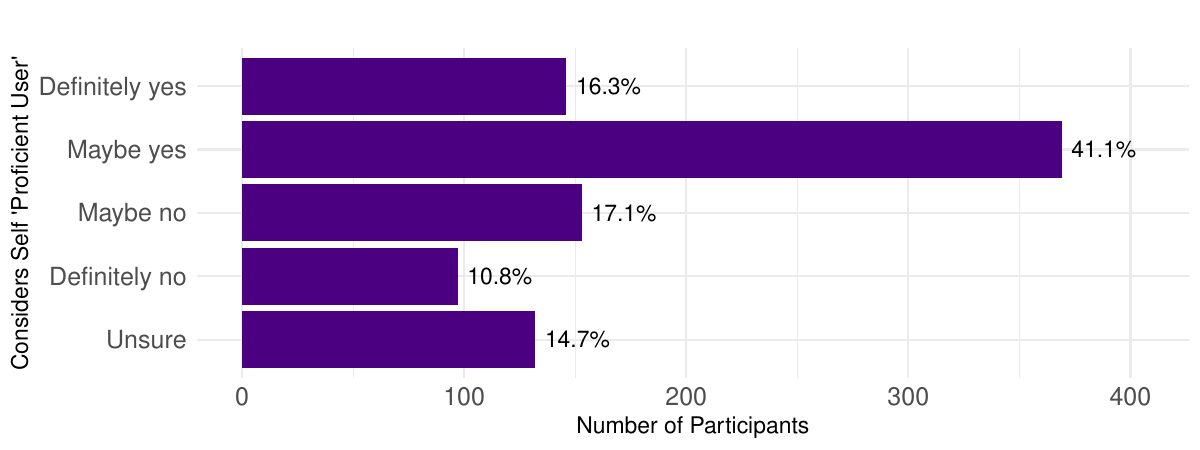}
        \caption{Participants' responses to ``Do you consider yourself a proficient user of GenAI?'' We define ``proficient user'' in the text.}
        \label{fig:proficient-user}
    \end{subfigure}
    \caption{Summary of user characteristics of the 897 GenAI users in our study sample. Percentage labels are computed as \% of these 897 users; \textit{GenAI Tools Used} and \textit{GenAI Use Cases} were multi-select characteristics, so totals will exceed 100\%.}
    \label{fig:user-characteristics}
    \Description{This figure contains four bar charts in a 2x2 grid summarizing characteristics of 897 generative AI users. Top left shows a horizontal bar chart of participants' reported frequency of GenAI use in the past 30 days, with bars extending rightward showing percentage of users for each frequency category from least to most frequent usage. Top right displays a horizontal bar chart of participants' reported GenAI use cases, showing various applications like writing assistance, creative tasks, coding help, and research, with bar lengths indicating popularity; this was multi-select so percentages exceed 100\%. Bottom left presents a horizontal bar chart of participants' reported GenAI tools used, listing various AI platforms and tools with bars showing percentage of users for each tool; also multi-select with totals exceeding 100\%. Bottom right shows a horizontal bar chart of participants' responses to "Do you consider yourself a proficient user of GenAI?" with bars indicating percentage selecting each response level on the assessment scale. All percentage labels represent proportions of the 897 GenAI users in the sample.
    }
\end{figure*}

Table ~\ref{tab:participant-demographics} provides a demographic summary of our sample, split into GenAI users 
(93\% of our participants) and GenAI non-users 
(6.5\% of our participants). 
Details on the usage characteristics of the 897 GenAI users in our study sample are available in Figure \ref{fig:user-characteristics}. 
\begin{table*}[h!]
\centering 
{
\begin{adjustbox}{max width=\textwidth}
{\small
\begin{tabular}{c | l |l r|l r}
\toprule
\multicolumn{2}{c}{} & \multicolumn{2}{c}{\textbf{Users} ($n = 897$)} & \multicolumn{2}{c}{\textbf{Non-users} ($n = 63$)} \\ \midrule

\multirow{2}{*}{\textbf{Sex}} 
& Female & 456 & 50.8\% & 35 & 55.6\% \\
& Male & 441 & 49.2\% & 28 & 44.4\% \\

\midrule

\multirow{6}{*}{\textbf{Age (Years)}} 
& 18-24 & 100 & 11.1\% & 9 & 14.3\% \\
& 25-34 & 162 & 18.1\% & 10 & 15.9\% \\
& 35-44 & 158 & 17.6\% & 8 & 12.7\% \\
& 45-54 & 145 & 16.2\% & 9 & 14.3\% \\
& 55-64 & 225 & 25.1\% & 19 & 30.2\% \\
& 65+ & 107 & 11.9\% & 8 & 12.7\% \\

\midrule

\multirow{5}{*}{\textbf{Race/Ethnicity}} 
& Asian & 54 & 6.0\% & 5 & 7.9\% \\
& Black & 108 & 12.0\% & 4 & 6.3\% \\
& Mixed & 97 & 10.8\% & 9 & 14.3\% \\
& White & 572 & 63.8\% & 40 & 63.5\% \\
& Other & 66 & 7.4\% & 5 & 7.9\% \\

\midrule

\multirow{3}{*}{\textbf{Tech Background}} 
& Yes & 262 & 29.2\% & 12 & 19.0\% \\
& No & 624 & 69.6\% & 50 & 81.0\% \\
& Decline to answer & 11 & 1.2\% & 0 & 0\% \\

\bottomrule
\end{tabular}}
\end{adjustbox}
}
\caption[GenAI User Study: Summary of participant demographics]{Summary of participant demographics, split into participants who reported using at least one GenAI tool (``Users'') and those who reported not using any GenAI tools (``Non-users''). Information on participants' sex, age, and race/ethnicity were provided by Prolific; data on participants' technical backgrounds were collected through our survey.}
\Description{
A demographic summary table comparing 897 GenAI users with 63 non-users across four categories. For sex: Users are 50.8\% female and 49.2\% male; non-users are 55.6\% female and 44.4\% male. For age: Both groups show similar distributions across age ranges from 18-24 to 65+, with the largest groups being 55-64 years (25.1\% users, 30.2\% non-users). For race/ethnicity: Both groups are predominantly White (63.8\% users, 63.5\% non-users), followed by Black (12.0\% users, 6.3\% non-users), Mixed (10.8\% users, 14.3\% non-users), Asian (6.0\% users, 7.9\% non-users), and Other (7.4\% users, 7.9\% non-users). For technical background: 29.2\% of users have technical backgrounds compared to 19.0\% of non-users, with most participants lacking technical backgrounds (69.6\% users, 81.0\% non-users).
}
\label{tab:participant-demographics}
\end{table*}

\section{Supplemental Results}

\subsection{Revisions to Preliminary Codebooks} \label{app:codebook-revisions}

Throughout analysis of participants' responses, we made four major revisions to our codebook of failure modes (we used Li et al.'s taxonomy of failure modes \cite{mapping_2025} as a starting point). First, we generalized improper use to \textit{Over-reliance}, for which we adopt Abercrombie et al.'s definition, ``Unfettered and/or obsessive belief in the accuracy or other quality of a [GenAI] system, resulting in complacency, lack of critical thinking and other actual or potential negative impacts'' \cite{abercrombie}. Second, we replaced failure to comply with contextual norms with two more granular failure modes: \textit{Poor/unsatisfactory quality output}, which we define as ``when GenAI fails to meet user expectations by performing poorly in some other way; e.g., producing gibberish, sycophancy'' and \textit{Parroting}, which we define as ``when GenAI `regurgitates' source material, possibly resulting in copyright infringement and/or privacy violations.'' Finally, we introduced \textit{Erroneous prompting}, which we define as ``When the user intends to engage in a plausible use case of GenAI, but provides prompts that are underspecified or incorrect for the task.'' See our full codebook of failure modes in Section~\ref{app:failure-modes}.

Based on analysis of participants' responses, we modified or introduced four risks to our preliminary codebook (we used Li et al.'s taxonomy of harms \cite{mapping_2025} as a starting point). First, we introduced \textit{Academic repercussions} and \textit{Workplace repercussions}, which we define as reprimand (i.e., from an instructor or employer) for the unsanctioned use of GenAI, possibly resulting in material consequences. These comprised a new risk category, \textit{Repercussions for unsanctioned use}. Second, we introduced \textit{Productivity loss}, which we define as ``End user's loss of productivity due to the underperformance of a GenAI application, including producing nonsensical or poor quality outputs that degrade its utility.'' Finally, we generalized \textit{Erosion of due process} to \textit{Legal consequences}, which we define broadly as ''Legal consequences as a result of use or misuse of GenAI.'' 

\subsection{Supplemental Tables and Figures}

\begin{figure*}[t]
    \centering
    \begin{subfigure}[t]{0.495\textwidth}
        \centering
        \includegraphics[width=.99\textwidth]{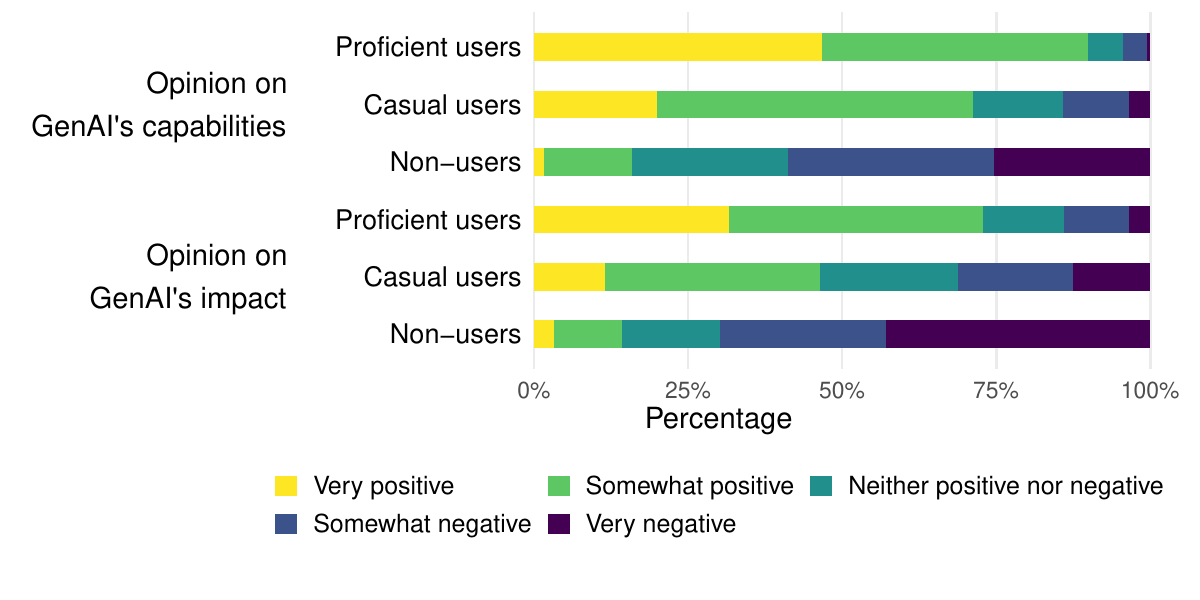}
        \caption{Participants' reported opinions on GenAI. In general, participants tend to have less positive opinions about GenAI's impact than its capabilities, but users tend to have positive opinions about both while non-users tend to have negative opinions about both. Proficient users had even more positive opinions than casual users.}
        \label{fig:opinions-comparison}
    \end{subfigure}\hfill
    \begin{subfigure}[t]{0.495\textwidth}
        \centering
        \includegraphics[width=.99\textwidth]{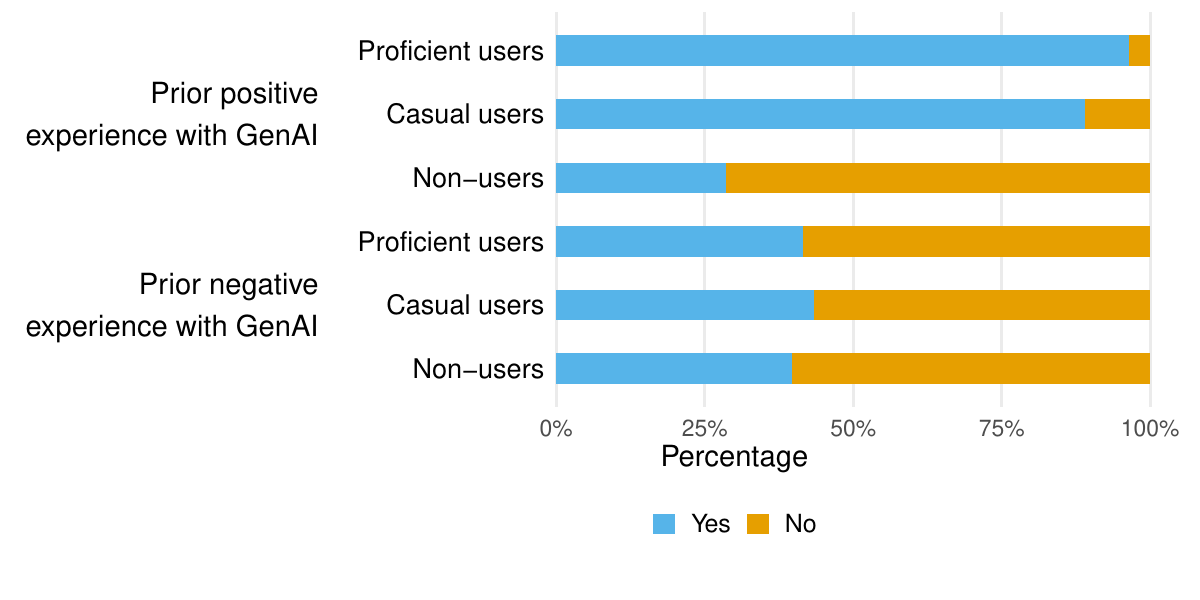}
        \caption{Participants' reported experiences with GenAI. Most users recalled a prior positive experience with GenAI, but most did not recall any prior negative experiences. Most non-users did not recall either a positive or negative experience with GenAI but were more likely to recall a negative one than a positive one.}
        \label{fig:experiences-comparison}
    \end{subfigure}
    \caption{Participants' reported opinions on and prior experiences with GenAI. Participants are split into three groups: Non-users, Casual users, and Proficient users.}
    \label{fig:opinions-and-experiences}
    \Description{
This figure contains two subfigures showing participant opinions and experiences with GenAI, split into three groups: Non-users, Casual users, and Proficient users.
(a) A stacked horizontal bar chart showing participants' opinions on GenAI's capabilities versus impact. For capabilities: Proficient users show approximately 90\% positive opinions (very positive and somewhat positive combined), casual users show about 75\% positive, while non-users show majority negative opinions. For impact: Proficient users show about 75\% positive opinions, casual users about 50\% positive, and non-users show majority negative opinions. The remaining responses are distributed across neutral, somewhat negative, and very negative categories.
(b) A stacked horizontal bar chart showing prior experiences with GenAI. For positive experiences: Most proficient and casual users report "Yes" while most non-users report "No". For negative experiences: Most users across all categories report "No", but non-users are slightly more likely to report negative experiences than positive ones.
}
\end{figure*}

\begin{figure*}[t]
    \centering
        \includegraphics[width=.7\textwidth]{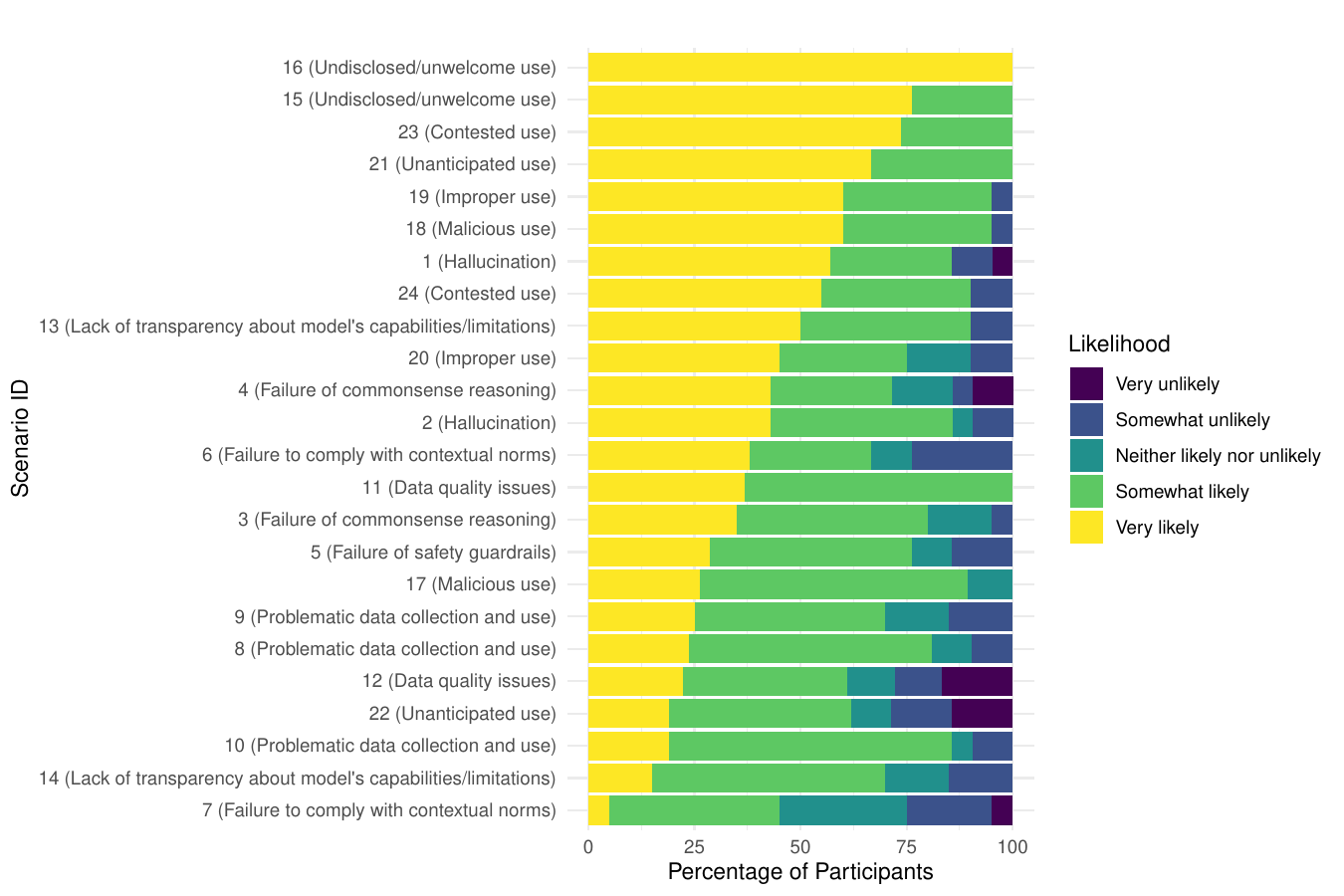}
        \caption{Participants' perceived likelihoods of each of our use-case-failure scenarios, with scenarios along the y-axis sorted from most to least frequently perceived as ``Very likely.'' Most participants thought most of our scenarios were at least somewhat likely. The scenarios representing the Undisclosed/unwelcome use failure mode were the most frequently perceived as Very likely.}
        \label{fig:ucfm-likelihoods}
    \Description{
A horizontal bar chart showing participants' perceived likelihoods of 24 different use-case-failure scenarios, with scenarios listed along the y-axis sorted from most to least frequently perceived as "Very likely". Each bar is divided into five colored segments representing likelihood ratings: Very likely, Somewhat likely, Neither likely nor unlikely, Somewhat unlikely, and Very unlikely. The scenarios representing Undisclosed/unwelcome use failure modes (scenarios 15 and 16) appear at the top as most frequently perceived as "Very likely", while scenario 7 (Failure to comply with contextual norms) appears at the bottom as least likely to be rated "Very likely". Most participants rated most scenarios as at least somewhat likely to occur.
}
\end{figure*}

\begin{table*}[t]
\centering 
{
\begin{adjustbox}{max width=\textwidth}
\begin{tabular}{c | l  r | c | c | l  r}
\toprule
& \multicolumn{2}{c|}{\textbf{Use-case-only}} & \multicolumn{4}{c}{\textbf{Use-case-failure}} \\ \midrule
\textbf{Scenario} & \textbf{Frequently Surfaced Risk(s)} & \textbf{$n$} &  \textbf{Failure Mode} & \textbf{Real-World Harm(s)} & \textbf{Frequently Surfaced Risk(s)} & \textbf{$n$} \\ \midrule

\multirow{2}{*}{1} & Academic repercussions & 12  
& \multirow{4}{*}{Hallucination} & Academic repercussions & Academic repercussions & 16  \\
& Propagating misconceptions/false beliefs & 4  & & & \\ \cline{1-3} \cline{5-7}

\multirow{2}{*}{2} & Legal repercussions & 14  
& & Legal repercussions & Legal repercussions & 17  \\
& Financial/earnings loss & 6  & & & & \\ \midrule

\multirow{2}{*}{3} & Productivity loss & 13  
& \multirow{4}{*}{\shortstack{Failure of\\commonsense\\reasoning}} & Productivity loss & Propagating misconceptions/false beliefs & 7  \\
& Propagating misconceptions/false beliefs & 9 &  & & Loss of confidence/trust & 4  \\ \cline{1-3} \cline{5-7}

\multirow{2}{*}{4} & Bodily injury & 14  
& & Bodily injury & Bodily injury & 17  \\
& Loss of life & 13 &  & & & \\ \midrule

\multirow{2}{*}{5} & Loss of confidence/trust & 10  
& \multirow{2}{*}{\shortstack{Failure of\\safety guardrails}} & Productivity loss & Productivity loss & 8  \\
& Toxic content & 8 &  & & & \\ \midrule

\multirow{2}{*}{6} & Productivity loss & 10  
& \multirow{4}{*}{\shortstack{Failure to comply\\with\\contextual norms}} & Productivity loss & Productivity loss & 7  \\
& Academic repercussions & 6 &  & & Academic repercussions & 6  \\ \cline{1-3} \cline{5-7}

\multirow{2}{*}{7} & Academic repercussions & 15  
&  & Academic repercussions & Productivity loss & 12  \\
& Loss of creativity/critical thinking & 5 &  & & & \\ \midrule

\multirow{2}{*}{8} & {\cellcolor[gray]{0.85}Academic repercussions} & 11  
& \multirow{5}{*}{\shortstack{Problematic data\\collection and use}} & IP/copyright/personality rights loss & Financial/earnings loss & 8  \\
& {\cellcolor[gray]{0.85}Propagating misconceptions/false beliefs} & 4 &  & & & \\ \cline{1-3} \cline{5-7}

\multirow{2}{*}{9} & {\cellcolor[gray]{0.85}
Productivity loss} & 8  
&  & Disclosure & Disclosure & 10  \\
& {\cellcolor[gray]{0.85}Bodily injury} & 4 &  & & & \\ \cline{1-3} \cline{5-7}

\multirow{1}{*}{10} & {\cellcolor[gray]{0.85}Productivity loss} & 18  
& & Exclusion & Exclusion & 7   \\ \midrule

\multirow{2}{*}{11} & {\cellcolor[gray]{0.85}IP/copyright/personality rights loss} & 9  
& \multirow{4}{*}{\shortstack{Data quality\\issues}} & Stereotyping & Stereotyping & 6  \\
& {\cellcolor[gray]{0.85}Productivity loss} & 9 &  & Academic repercussions & &  \\ \cline{1-3} \cline{5-7}

\multirow{2}{*}{12} & Workplace repercussions & 7  
& & Stereotyping & Stereotyping & 6   \\
& & & & Workplace repercussions & &  \\ \midrule

\multirow{2}{*}{13} & Financial/earnings loss & 15  
& \multirow{4}{*}{\shortstack{Lack of\\transparency about\\ model's capabilities/\\limitations}} & Legal repercussions & Legal repercussions & 9  \\
& Legal repercussions & 12 &  & & &  \\ \cline{1-3} \cline{5-7}

\multirow{2}{*}{14} & Loss of life & 12  
& & Loss of life & Loss of life & 12   \\
& Bodily injury & 10 &  & & &  \\ \midrule

\multirow{2}{*}{15} & Loss of confidence/trust & 9  
& \multirow{4}{*}{\shortstack{Undisclosed/\\unwelcome use}} & Overburdening ecosystems & Loss of creativity/critical thinking & 4  \\
&  &  &   & & Overburdening ecosystems & 4  \\ \cline{1-3} \cline{5-7}

\multirow{2}{*}{16} & Loss of creativity/critical thinking & 15  
& & Overburdening ecosystems & Academic repercussions & 10  \\
& Academic repercussions & 11 &  & Academic repercussions & & \\ \midrule

\multirow{2}{*}{17} & {\cellcolor[gray]{0.85}Loss of creativity/critical thinking} & 10 
& \multirow{4}{*}{Malicious use} & Propagating misconceptions/false beliefs & Propagating misconceptions/false beliefs & 7   \\
&  {\cellcolor[gray]{0.85}Academic repercussion}s & 4 &   &  & & \\ \cline{1-3} \cline{5-7}

\multirow{2}{*}{18} & Propagating misconceptions/false beliefs & 8  
& & Pollution of information ecosystem & Propagating misconceptions/false beliefs & 12  \\
& Loss of confidence/trust & 7 &  & & & \\ \midrule

\multirow{2}{*}{19} & Legal repercussions & 12 
& \multirow{4}{*}{Improper use} & Legal repercussions & Legal repercussions & 9   \\
&  Financial/earnings loss & 7 &   & Loss of confidence/trust & & \\ \cline{1-3} \cline{5-7}

\multirow{2}{*}{20} & Breach of ethics/values/norms & 6  
& & Breach of ethics/values/norms & Breach of ethics/values/norms & 10  \\
& Academic repercussions & 5 &  & Productivity loss & & \\ \midrule

\multirow{2}{*}{21} & Academic repercussions & 15 
& \multirow{4}{*}{Unanticipated use} & Breach of ethics/values/norms & Academic repercussions & 17   \\
&  Loss of creativity/critical thinking & 7 &   & & & \\ \cline{1-3} \cline{5-7}

\multirow{2}{*}{22} & Emotional distress & 6 &  & Toxic content & Legal repercussions & 4 \\
& Toxic content & 6 &  & & Toxic content & 4  \\ \midrule

\multirow{2}{*}{23} & Defamation/libel/slander & 8 
& \multirow{4}{*}{Contested use} & Defamation/libel/slander & Defamation/libel/slander & 12   \\
&  &  & & Erosion of trust in public information & & \\ \cline{1-3} \cline{5-7}

\multirow{2}{*}{24} & Overburdening ecosystems & 12  
& & Opportunity loss & Opportunity loss & 10  \\
& Loss of confidence/trust & 7 &  & Loss of confidence/trust & &  \\ 

\bottomrule
\end{tabular}
\end{adjustbox}
}
\caption[GenAI User Study: Summary of risks surfaced within each use-case-failure scenario]{Summary of participants' frequently described risks within each scenario, with the use-case-only format on the left and the use-case-failure format on the right. For the use-case-failure format, we present participants' frequently surfaced risk(s) alongside the real-world harm that was either associated with the incident that the given scenario was based on or discussed as reasonably likely by the research team. We include all risks that were described by $\geq 20\%$ of participants within each study condition (each condition had between 18 and 21 participants). We highlight in gray where participants' frequently surfaced risks were disjoint from the associated real-world harms.}
\Description{
A table summarizing the frequently surfaced risks for each of the 24 scenarios, comparing use-case-only format (left side) with use-case-failure format (right side). The table has columns for Scenario number, Frequently Surfaced Risk(s), participant count (n), Failure Mode, Real-World Harm(s), and corresponding data for the use-case-failure format. For each scenario, risks that were described by ≥20\% of participants are included. The table shows various risk types including Academic repercussions, Legal repercussions, Productivity loss, Bodily injury, Loss of life, Financial/earnings loss, Propagating misconceptions/false beliefs, and others. Gray highlighting indicates where participants' frequently surfaced risks were disjoint from the associated real-world harms, suggesting gaps in risk perception.
}
\label{tab:risks-by-scenario}
\end{table*}
\clearpage


\newpage
\section{Codebook of Failure Modes}
\label{app:failure-modes}






{\footnotesize
\linespread{1.3}\selectfont
\begin{tabularx}{\textwidth}{X|X}

\hline {\cellcolor[gray]{0.85}\textbf{Failure Mode}} & {\cellcolor[gray]{0.85}\textbf{Description}} \\ \hline 

Hallucination & Making false, misleading, or inaccurate claims as if they were facts \\ \hline

Failure of commonsense reasoning & Generative systems do not have intrinsic logic/commonsense reasoning \\ \hline

Failure of safety guardrails & Safety feature meant to prevent problematic outputs either produces a new problem or just fails \\ \hline

Parroting & When Generative AI "regurgitates" source material. Parroting may result in outright copyright infringement and/or privacy violations. \\ \hline

Poor/unsatisfactory quality output & AI fails to meet user expectations by performing poorly in some other way; e.g., "AI can make mistakes" or producing gibberish \\ \hline

Problems with data collection and processing & The indiscriminate nature of AI developers' collection and/or use of data (especially internet data), preventing users from giving informed consent on how their data is used. \\ \hline

Data quality issues & Data reflects or exacerbates societal biases, or overrepresents a certain type of content, or is polluted with poor quality information which is not sufficiently cleaned. \\ \hline

Lack of transparency about model's capabilities/limitations & Developer fails to provide clarity about what their system can and/or should be used for, possibly resulting in unintended use \\ \hline

Undisclosed or unwelcome use & Use in a context where there is an expectation of human expertise, labor, or creativity \\ \hline

Contested use & It is unclear whether or not GenAI was used in a certain context, possibly resulting in false accusations of dishonesty \\ \hline

Over-reliance & Unfettered and/or obsessive belief in the accuracy or other quality of a technology system, resulting in complacency, lack of critical thinking and other actual or potential negative impacts \\ \hline

Malicious use & Bad actor uses generative AI to facilitate disinformation, an influence operation, fraud, defamation, nonconsensual sexualization, or security threat \\ \hline

Erroneous prompting & User misunderstands some aspect of how to get what they want out of the system, but what they want does fall within intended use \\ \hline

Lack of robustness against adversarial prompting & Model misbehaves as a result of adversarial prompting \\ \hline

Resource demands & Training and using genAI for inference are costly in terms of both physical and human resources \\ \hline

Unanticipated use & User uses the system for a purpose that was unanticipated and/or outside of the scope of intended uses (as conceived of by the developer/deployer) \\ \hline

Improper use & Using or deploying GenAI to complete tasks either requiring professional license/training or subject to industry standards, and failing to properly review outputs. \\ \hline

Generic AI failure & AI failure that does not fall into the above categories; e.g., issues of alignment \\ \hline

N/A & Misunderstood our question or said nothing could go wrong \\ \hline

\end{tabularx}
}
\newpage
\section{Codebook of Risks}
\label{app:risk-defs}

{\footnotesize
\linespread{1.2}\selectfont

\begin{xltabular}
{\textwidth}
{ X | X }
\label{tab:harm-modes}\\

\hline \textbf{Harm} & \textbf{Description} \\ \hline 
\endfirsthead

\multicolumn{2}{c}%
{\tablename\ \thetable{} -- continued from previous page} \\
\hline \textbf{Risk} & \textbf{Description} \\ \hline 
\endhead

\hline \multicolumn{2}{|r|}{{Continued on next page}} \\ \hline
\endfoot

\hline
\endlastfoot

\hline \textbf{Risk} & \textbf{Description} \\ \hline
\endfirsthead

\endhead

\endfoot

\endlastfoot

{\cellcolor[gray]{0.85}\textbf{\textit{Representation \& Toxicity Harms}}} & {\cellcolor[gray]{0.85} \textbf{\textit{AI systems under-, over-, or misrepresenting certain groups or generating toxic, offensive, abusive, or hateful content}}} \\ \hline
Cultural dispossession & Intentional and/or unintentional erasure of cultural goods and values, such as ways of speaking, expressing humour, or sounds and voices that contribute to a cultural identity, or their inappropriate re-use in other cultures \\ \hline
Stereotyping & Derogatory or otherwise harmful stereotyping or homogenisation of individuals, groups, societies or cultures due to the mis-representation, over-representation, under-representation, or non-representation of specific identities, groups or perspectives \\ \hline
Unfair capability distribution & Performing worse for some groups than others in a way that harms the worse-off group \\ \hline
Toxic content & Generating content that violates community standards, including harming or inciting hatred or violence against groups (e.g. gore, sexual content of children, profanities, identity attacks) \\ \hline

{\cellcolor[gray]{0.85} \textbf{\textit{Informational Harms}}} & {\cellcolor[gray]{0.85} \textbf{\textit{AI systems generating and facilitating the spread of inaccurate or misleading information that causes people to develop false beliefs}}} \\ \hline
Propagating misconceptions/false beliefs & Generating or spreading false, low-quality, misleading, or inaccurate information that causes people to develop false or inaccurate perceptions and beliefs \\ \hline
Erosion of trust in public information & Eroding trust in public information and knowledge \\ \hline
Pollution of information ecosystem & Contaminating publicly available information with false or inaccurate information (i.e., the generative tool's output is disseminated beyond the end user) \\ \hline

{\cellcolor[gray]{0.85}\textbf{\textit{Autonomy}}} & {\cellcolor[gray]{0.85}\textbf{\textit{Loss of or restrictions to the ability or rights of an individual, group or entity to make decisions and control their identity and/or output due to the use of misuse of a technology system or set of systems}}} \\ \hline
Autonomy/agency loss & Loss of an individual, group or organisation's ability to make informed decisions or pursue goals \\ \hline
Impersonation/identity theft & Theft of an individual, group or organisation's identity by a third-party in order to defraud, mock or otherwise harm them or another party \\ \hline
IP/copyright/personality rights loss & Misuse or abuse of an individual or organisation's intellectual property, including copyright, trademarks, and patents \& Loss of or restrictions to the rights of an individual to control the commercial use of their identity, such as name, image, likeness, or other unequivocal identifiers \\ \hline

{\cellcolor[gray]{0.85}\textbf{\textit{Physical}}} & {\cellcolor[gray]{0.85}\textbf{\textit{Physical injury to an individual or group, or damage to physical property due to the use of misuse of a technology system or set of systems}}} \\ \hline
Bodily injury & Physical pain, injury, illness, or disease suffered by an individual or animal or group due to the malfunction, use or misuse of a technology system \\ \hline
Self-harm & A person who deliberately damages their own body as a direct or indirect result of using a technology system \\ \hline
Loss of life & Accidental or deliberate loss of life, including suicide, extinction or cessation, due to the use or misuse of a technology system \\ \hline
Personal health deterioration & Physical deterioration of an individual or animal over time in the form of disease, organ failure, prolonged hospital stay or death, etc \\ \hline
Property damage & Action(s) that lead directly or indirectly to the damage or destruction of tangible property eg. buildings, possessions, vehicles, robots \\ \hline

{\cellcolor[gray]{0.85}\textbf{\textit{Psychological}}} & {\cellcolor[gray]{0.85}\textbf{\textit{Impairment of the psychological mental health and wellbeing of an individual, group or organisation due to the use of misuse of a technology system or set of systems}}} \\ \hline
Addiction & Emotional or material dependence on technology or a technology system \\ \hline
Alienation/isolation & An individual's or group's feeling of lack of connection with those around as a result of technology system use or misuse \\ \hline
Emotional distress & Distress, possibly severe and lasting, as a result of use or misuse of a generative system \\ \hline
Coercion/manipulation & Use of a technology system to covertly alter user beliefs and behaviour using nudging, dark patterns and/or other opaque techniques \\ \hline
Dehumanization/objectification & Use or misuse of a technology system to depict and/or treat people as not human, less than human, or as objects \\ \hline
Harassment/abuse/intimidation & Online behaviour such as sexual harassment that makes an individual or group feel alarmed or threatened \\ \hline
Over-reliance & Unfettered and/or obsessive belief in the accuracy or other quality of a technology system, resulting in complacency, lack of critical thinking and other actual or potential negative impacts \\ \hline
Radicalization & Adoption of extreme political, social or religious ideals and aspirations due to the nature, use or misuse of an algorithmic system \\ \hline
Sexualization & The non-consensual sexualisation of an individual or group using a technology or application \\ \hline

{\cellcolor[gray]{0.85}\textbf{\textit{Reputational}}} & {\cellcolor[gray]{0.85}\textbf{\textit{Damage to the reputation of an individual, group or organisation due to the use of misuse of a technology system or set of systems}}} \\ \hline
Defamation/libel/slander & Use of a technology system to create, facilitate or amplify false perception(s) about an individual, group or organisation \\ \hline
Loss of confidence/trust & The use or misuse of a technology system that leads directly or indirectly to the loss of confidence or trust in either the end user or the developer/deployer \\ \hline

{\cellcolor[gray]{0.85}\textbf{\textit{Financial \& Business}}} & {\cellcolor[gray]{0.85}\textbf{\textit{Damage to the financial interests of an individual or group, or to the strategic, operational, legal or financial interests of a business due to the use of misuse of a technology system or set of systems}}} \\ \hline
Business operations/infrastructure damage & Damage, disruption or destruction of a business system and/or its components \\ \hline
Confidentiality loss & Unauthorised sharing of sensitive, confidential information and documents such as corporate strategy and financial plans with third-parties, risking loss of market position or revenue \\ \hline
Financial/earnings loss & Loss of money, income or value due to the use, misuse, or underperformance of a genAI application \\ \hline
Livelihood loss & An individual or group's loss of ability to support themselves financially or vocationally due to etc, resulting in inability to buy food, reduced employment prospects, bankruptcy, foreclosure, homelessness, etc \\ \hline
Increased competition & Enhanced competition due to the inappropriate or unethical use or misuse of a technology system to gain market share \\ \hline
Monopolization & Abuse of market power through the control of prices, thereby limiting competition and creating unfair barriers to entry \\ \hline
Opportunity loss & Loss of ability to take advantage of a financial or other opportunity, such as education, immigration, employability/securing a job \\ \hline

{\cellcolor[gray]{0.85}\textbf{\textit{Human Rights \& Civil Liberties}}} & {\cellcolor[gray]{0.85}\textbf{\textit{Use or misuse of a technology system in a manner that compromises fundamental human rights and freedoms}}}  \\ \hline
Benefits/entitlements loss & Denial of or loss of access to welfare benefits, pensions, housing, etc due to the malfunction, use or misuse of a technology system \\ \hline
Dignity loss & Perceived loss of value experienced by or disrespect shown to an individual or group, resulting in self-sheltering, loss of connections and relationships, and public stigmatization \\ \hline
Discrimination & Unfair or inadequate treatment or arbitrary distinction based on a person's race, ethnicity, age, gender, sexual preference, religion, national origin, marital status, disability, language, or other protected groups \\ \hline
Forced labor & Restrictions to and loss of people's right to be held in slavery or servitude, required to perform forced or compulsory labour, or trafficked in \\ \hline
Loss of freedom of speech/expression & Restrictions to or loss of people's right to articulate their opinions and ideas without fear of retaliation, censorship, or legal sanction \\ \hline
Loss of freedom of assembly/association & Restrictions to or loss of people's right to come together and collectively express, promote, pursue, and defend their collective or shared ideas, and/or to join an association \\ \hline
Loss of social rights and access to public services & Restrictions to or loss of rights to work, social security, and adequate standard of living, housing, health and education \\ \hline
Loss of right to information & Restrictions to or loss of people's right to seek, receive and impart information held by public bodies \\ \hline
Legal consequences & Legal consequences as a result of use or misuse of GenAI \\ \hline

{\cellcolor[gray]{0.85}\textbf{\textit{Societal \& Cultural}}} & 
{\cellcolor[gray]{0.85}\textbf{\textit{Harms affecting the functioning of societies, communities and economies caused directly or indirectly by the use or misuse of a technology system or set of systems}}} \\ \hline
Breach of ethics/values/norms & An actual or perceived violation or deviation from the established societal values, norms or ethical standards or principles \\ \hline
Overburdening ecosystems & Pollution of a space/ecosystem that is expected to be free of AI involvement/influence (e.g., creative material submission portals, job applications) \\ \hline
Cheating/plagiarism & Use of generative AI in an academic setting to either cheat or plagiarize \\ \hline
Chilling effect & The creation of a climate of self-censorship that deters democratic actors such as journalists, advocates and judges from speaking out \\ \hline
Damage to public health & Adverse impacts on the health of groups, communities or societies, including malnutrition, disease and infection conditions \\ \hline
Job loss/losses & Replacement/displacement of human jobs by a technology system or set of systems, leading to increased unemployment, inequality, reduced consumer spending and social friction \\ \hline
Labor exploitation & Use/misuse of labour to help train, develop, manage or optimise a technology system or set of systems, including under-paid and/or offshore \\ \hline
Loss of creativity/critical thinking & Devaluation and/or deterioration of human creativity, artistic expression, imagination, critical thinking or problem-solving skills \\ \hline
Public service delivery deterioration & Poor performance of a public technology system due to malfunction, over-use, under-staffing etc, resulting in individuals, groups, or organisations unable to use it in a manner they can reasonably expect \\ \hline
Societal destabilization & Societal instability in the form of strikes, demonstrations and other types of civil unrest caused by loss of jobs to technology, unfair algorithmic outcomes, disinformation, etc \\ \hline
Societal inequality & Increased difference in social status or wealth between individuals or groups caused or amplified by a technology system, leading to the loss of social and community wellbeing/cohesion and destabilisation \\ \hline

{\cellcolor[gray]{0.85}\textbf{\textit{Political \& Economic}}} & {\cellcolor[gray]{0.85}\textbf{\textit{Damage to core political and economic institutions and the effective delivery of government services caused by the use or misuse of a technology system or set of systems}}} \\ \hline
Economic manipulation & Generative AI facilitating targeted manipulation of public opinion for economic purposes (e.g., inflating stock prices) \\ \hline
Critical infrastructure damage & Damage, disruption to or destruction of systems essential to the functioning and safety of a nation or state, including energy, transport, health, finance and communication systems \\ \hline
Economic instability & Uncontrolled fluctuations impacting the financial system, or parts thereof, due to the use or misuse of a technology system, or set of systems \\ \hline
Power concentration & Amplification or concentration of economic and/or political wealth and power, resulting in increased inequality and instability \\ \hline
Institutional trust loss & Erosion of trust in public institutions and weakened checks and balances due to mis/disinformation, influence operations, or real or perceived misuse of generative AI \\ \hline
Political instability & Political unrest caused directly or indirectly by the use or misuse of a technology system \\ \hline
Political manipulation or interference & Manipulation of the beliefs and behaviours of individuals or groups for political purposes, including dissemination of false, generated, and/or misleading information that can interrupt or mislead voters and/or undermine trust in electoral processes \\ \hline

{\cellcolor[gray]{0.85}\textbf{\textit{Environmental}}} & {\cellcolor[gray]{0.85}\textbf{\textit{Damage to the environment caused by the use or misuse of a technology system or set of systems}}} \\ \hline
Biodiversity loss & Over-expansion of technology infrastructure, or inadequate alignment of technology with sustainable practices, leading to deforestation, habitat destruction and the fragmentation and loss of biodiversity \\ \hline
Carbon emissions & Release of carbon dioxide, nitric oxide and other gases, increasing carbon emissions, exacerbating climate change, and negatively impacting local communities \\ \hline
Electronic waste & Electrical or electronic equipment that is waste, including all components, sub-assemblies and consumables that are part of the equipment at the time the equipment becomes waste \\ \hline
Excessive energy consumption & Excessive energy use resulting in energy bottlenecks and shortages for communities, organisations and businesses \\ \hline
Excessive water consumption & Excessive use of water to cool data centres and for other purposes, leading to water restrictions or shortages for local communities or businesses \\ \hline
Pollution & Actual or potential pollution to the air, ground, noise, or water caused by a technology system \\ \hline

{\cellcolor[gray]{0.85}\textbf{\textit{Privacy \& Security}}} & {\cellcolor[gray]{0.85}\textbf{\textit{AI systems leaking, reproducing, generating or inferring sensitive, private, hazardous, or secured information}}}  \\ \hline
Secondary use & The use of personal data collected for one purpose for a different purpose without end-user consent; AI exacerbates secondary use risks by creating new AI capabilities with collected personal data, and (re)creating models from a public dataset \\ \hline
Exclusion & The failure to provide end-users with notice and control over how their data is being used; AI exacerbates exclusion risks by training on rich personal data without consent \\ \hline
Exposure & Revealing sensitive private information that people view as deeply primordial that we have been socialized into concealing; AI creates new types of exposure risks through generative techniques that can reconstruct censored or redacted content; and through exposing inferred sensitive data, preferences, and intentions \\ \hline
Disclosure & Revealing and improperly sharing data of individuals; AI creates new types of disclosure risks by inferring additional information beyond what is explicitly captured in the raw data; AI exacerbates disclosure risks through sharing personal data to train models \\ \hline
Distortion & Disseminating false or misleading information about people \\ \hline
Insecurity & Carelessness in protecting collected personal data from leaks and improper access due to faulty data storage and data practices \\ \hline
Dissemination of dangerous information & Leaking, generating or correctly inferring hazardous or sensitive information that could pose a security threat \\ \hline
Cyberattacks & Generative AI facilitating the damage, disruption or destruction of a third-party system and/or its components via malfunction, cyberattacks, etc \\ \hline
Surveillance & Watching, listening to, or recording of an individual's activities \\ \hline

{\cellcolor[gray]{0.85}\textbf{\textit{Repercussions for Unsanctioned Use}}} & {\cellcolor[gray]{0.85}\textbf{\textit{Punishment for the unsanctioned use of AI}}} \\ \hline
Academic repercussions & Reprimand from an instructor for unsanctioned use of AI (i.e. to cheat or plagiarize), possibly resulting in a poor grade or other punishment \\ \hline
Workplace repercussions & Reprimand from an employer for unsanctioned use of AI, possibly resulting in termination of employment or other punishment \\ \hline

\multicolumn{2}{c}{\cellcolor[gray]{0.85}\textbf{\textit{Other}}} \\ \hline
Productivity loss & End user's loss of productivity due to the underperformance of a genAI application, including producing nonsensical or poor quality outputs, degrading its utility \\ \hline

\end{xltabular}

}
\newpage
\section{Codebook of Stakeholders}
\label{app:stakeholders}

{\footnotesize
\linespread{1.3}\selectfont
\begin{tabularx}{\textwidth}{X|X}
\hline {\cellcolor[gray]{0.85}\textbf{Stakeholder}} & {\cellcolor[gray]{0.85}\textbf{Description}} \\ \hline
General public & Members of society \\ \hline
Advocacy groups & NGOs or civil society organizations that represent affected communities \\ \hline
AI corporations & Companies that develop, deploy, or commercialize AI systems \\ \hline
Regulators and Government Bodies & Courts, regulators, and agencies that investigate or impose consequences \\ \hline
Media & Journalists and news outlets \\ \hline
End users & Individuals or organizations that use AI systems \\ \hline
Academic institutions & Universities, colleges, or other institutions for educational purpose \\ \hline
Researchers in general & research scientists \\ \hline
Auditors and Oversight Boards & Entities that assess AI risks or harms through audits or review \\ \hline
N/A & not answering the question \\ \hline
Unsure & people saying "no idea" \\ \hline
\end{tabularx}
}
\newpage
\section{Codebook of Opinions}
\label{app:opinions}

{\footnotesize
\linespread{1.2}\selectfont

\begin{xltabular}
{\textwidth}
{ X | X }
\label{tab:failure-modes}\\

\hline {\cellcolor[gray]{0.85}\textbf{Code}} & {\cellcolor[gray]{0.85}\textbf{Description}} \\ \hline 
\endfirsthead

\multicolumn{2}{c}%
{\tablename\ \thetable{} -- continued from previous page} \\
\hline {\cellcolor[gray]{0.85}\textbf{Code}} & {\cellcolor[gray]{0.85}\textbf{Description}} \\ \hline 
\endhead

\hline \multicolumn{2}{|r|}{{Continued on next page}} \\ \hline
\endfoot

\hline
\endlastfoot
\hline {\cellcolor[gray]{0.85}\textbf{Code}} & {\cellcolor[gray]{0.85}\textbf{Description}} \\ \hline \endfirsthead

Hallucination & Making false, misleading, or inaccurate claims as if they were facts \\ \hline

Failure of commonsense reasoning & Generative systems do not have intrinsic logic/commonsense reasoning \\ \hline

Failure of safety guardrails & Safety feature meant to prevent problematic outputs either produces a new problem or just fails \\ \hline

Parroting & When Generative AI "regurgitates" source material. Parroting may result in outright copyright infringement and/or privacy violations. \\ \hline

Poor/unsatisfactory quality output & AI fails to meet user expectations by performing poorly in some other way; e.g., "AI can make mistakes" or producing gibberish \\ \hline

Problems with data collection and processing & The indiscriminate nature of AI developers' collection and/or use of data (especially internet data), preventing users from giving informed consent on how their data is used. \\ \hline

Data quality issues & Data reflects or exacerbates societal biases, or overrepresents a certain type of content, or is polluted with poor quality information which is not sufficiently cleaned. \\ \hline

Lack of transparency about model's capabilities/limitations & Developer fails to provide clarity about what their system can and/or should be used for, possibly resulting in unintended use \\ \hline

Undisclosed or unwelcome use & Use in a context where there is an expectation of human expertise, labor, or creativity \\ \hline

Contested use & It is unclear whether or not GenAI was used in a certain context, possibly resulting in false accusations of dishonesty \\ \hline

Unanticipated use & User uses the system for a purpose that was unanticipated and/or outside of the scope of intended uses (as conceived of by the developer/deployer) \\ \hline

Improper use & Using or deploying GenAI to complete tasks either requiring professional license/training or subject to industry standards, and failing to properly review outputs. \\ \hline

Malicious use & Bad actor uses generative AI to facilitate disinformation, an influence operation, fraud, defamation, nonconsensual sexualization, or security threat \\ \hline

Erroneous prompting & User misunderstands some aspect of how to get what they want out of the system, but what they want does fall within intended use \\ \hline

Lack of robustness against adversarial prompting & Model misbehaves as a result of adversarial prompting \\ \hline

Over-reliance & Unfettered and/or obsessive belief in the accuracy or other quality of a technology system, resulting in complacency, lack of critical thinking and other actual or potential negative impacts \\ \hline

Job loss/losses & Replacement/displacement of human jobs by a technology system or set of systems, leading to increased unemployment, inequality, reduced consumer spending and social friction \\ \hline

Productivity loss & End user's loss of productivity due to the underperfomance of a genAI application, including producing nonsensical or poor quality outputs, degrading its utility. \\ \hline

Labor exploitation & Use/misuse of labour to help train, develop, manage or optimise a technology system or set of systems, including under-paid and/or offshore \\ \hline

Loss of creativity/critical thinking & Devaluation and/or deterioration of human creativity, artistic expression, imagination, critical thinking or problem-solving skills \\ \hline

Reinforcing biases & The use of Generative AI may reinforce people's existing biases by affirming rather than challenging them \\ \hline

Alienation/isolation & An individual's or group's feeling of lack of connection with those around as a result of technology system use or misuse \\ \hline

Environmental harm & Harm to the environment (e.g., excessive water consumption, excessive energy consumption) \\ \hline

Informational harm & Generic informational harm (i.e., "misinformation") with no further specification \\ \hline

Pollution of information ecosystem & Contaminating publicly available information with false or inaccurate information (i.e., the generative tool's output is disseminated beyond the end user) \\ \hline

Societal inequality & Societal instability in the form of strikes, demonstrations and other types of civil unrest caused by loss of jobs to technology, unfair algorithmic outcomes, disinformation, etc \\ \hline

Reckless development/adoption & Participant expresses a belief that Generative AI is being developed and/or integrated into daily life too quickly, which may have consequences for people and their institutions \\ \hline

Sycophancy & Participant expresses dissatisfaction at Generative AI's sycophantic tendencies \\ \hline

Prefers human-originated & Participant expresses a belief that Generative AI cannot replace humans and/or that human-made creations are better \\ \hline

Developer/user/use case dependent & Participant expresses a belief that the ultimate impact of Generative AI is dependent on its developers (i.e., "who controls it"), users, and/or the contexts in which it is used. For example, participant acknowledges that GenAI has certain limitations (e.g., hallucination) but suggests that it can be helpful anyway with users' fact checking, etc. \\ \hline 

Purely positive & Participant expressed only positive sentiments about Generative AI \\ \hline

Existential risk & substantial progress in artificial general intelligence (AGI) could lead to human extinction  \\ \hline

Unsure & Participant expresses ambivalence or a belief that the ultimate outcome of Generative AI is unpredictable \\ \hline

Generic negative sentiment & Participant expresses that they believe Generative AI is bad, limited, and/or can cause harm without enough detail to specify further \\ \hline

N/A & Misunderstood our question, irrelevant, or unintelligible \\

\end{xltabular}
}

\end{document}